\documentclass[twocolumn]{aastex631}
\usepackage{amsmath}

\shorttitle{The N-PDF of Cygnus-X}
\shortauthors{Xing et al.}

\newcommand{\figwidth}{\linewidth}

\begin{document}

\title{The Column Density Probability Density Function of Cygnus-X}

\author[0000-0002-3411-9654]{Yuchen Xing}
\affiliation{National Astronomical Observatories, Chinese Academy of Sciences, Beijing 100101, China}
\affiliation{University of Chinese Academy of Sciences, Beijing 100049, China}

\author[0000-0002-5093-5088]{Keping Qiu}
\affiliation{School of Astronomy and Space Science, Nanjing University, 163 Xianlin Avenue, Nanjing 210023, China}
\email{kpqiu@nju.edu.cn}
\affiliation{ Key Laboratory of Modern Astronomy and Astrophysics (Nanjing University), Ministry of Education, Nanjing 210023, China}

\begin{abstract}
The density distribution within molecular clouds offers critical insights into their underlying physical processes, which are essential for understanding star formation. As a statistical measure of column density on the cloud scale, the shape and evolution of the column density probability density function (N-PDF) serve as important tools for understanding the dynamics between turbulence and gravity. Here we investigate the N-PDFs of Cygnus-X using the column density map obtained from Herschel, supplemented by HI and Young Stellar Objects (YSO) data. We find that the N-PDFs of Cygnus-X and four sub-regions display log-normal + power-law shapes, indicating the combined effects of turbulence and gravity in sculpting the density structure. 
We find evidence that the power-law segment of the N-PDFs flattens over time, and the transitional column density can be seen as a unique and stable star formation threshold specific to each molecular cloud. These results not only clarify the physical state of Cygnus-X but also emphasize the utility of the N-PDF as a statistical diagnostic tool, as it is an accessible indicator of evolutionary stages and star formation thresholds in molecular clouds.
\end{abstract}

\section{Introduction}

Molecular clouds are intricate environments where the interplay between turbulence and gravity, as well as stellar feedback and magnetic fields, plays a crucial role in shaping the star formation process. Unraveling the physical state of these clouds is essential to understanding the processes that lead to star formation. The column density probability density function (N-PDF) has emerged as a widely used and effective tool in this endeavor, providing valuable insights into the density distribution of molecular clouds.

The N-PDF of a molecular cloud is defined as the probability of the column density falling in the range $[N, N+dN]$. In practice, the column density $N$ is usually expressed in logarithmic scale by defining $\eta=\ln(N/N_{\rm peak})$, where $N_{\rm peak}$ is the most probable value of the N-PDF\footnote{Note that $\eta=\ln(N/<N>)$, where $<N>$ is the average column density, is also commonly used. For further details on the relationship between these notations, see \citet{Ossenkopf-Okada2016}.}. Whether from simulations \citep{Ballesteros-Paredes2011,Collins2012,Ward2014,Burkhart2015,Kortgen2019} or observations \citep{Kainulainen2009,Pineda2010,Russeil2013,Lombardi2015,Schneider2015_1}, the most basic shape of a star-forming molecular cloud's N-PDF is log-normal + power-law. At low column densities, isothermal supersonic turbulence leads to a log-normal N-PDF \citep{Vazquez-Semadeni1994}. For the same reason, a pure log-normal N-PDF is likely to be seen from a molecular cloud with no star formation since the whole cloud is dominated by turbulence \citep{Kainulainen2009,Schneider2012,Schneider2013}. The width $\sigma_\eta$ of the log-normal N-PDF shows the state of the turbulent gas and is a function of turbulence forcing parameter $b$ and sonic Mach number $M_s$, $\sigma_{\eta}^{2}=A\times\ln(1+b^{2}M_s^{2})$ \citep{Federrath2008,Burkhart2012,Burkhart2017}, where the scaling parameter $A$ is found to have values around $0.1$ \citep{Burkhart2012}. The participation of self-gravity makes the N-PDF of a star-forming molecular cloud present a power-law shape at high column densities \citep{Ballesteros-Paredes2011,Kritsuk2011,Collins2012,Ward2014,Girichidis2014,Burkhart2017}. The index of the power-law N-PDF is related to the gases' averaged density profile. Assuming spherical symmetry, for a power-law N-PDF with index $s$, we can calculate that its corresponding density profile $\rho \propto r^{-\alpha}$ has $\alpha=1+2/s$ \citep{Kritsuk2011,Federrath2013}. 

Based on the log-normal + power-law shape, the N-PDF of a star-forming molecular cloud may also exhibit additional components. Magnetic fields oriented perpendicular to the column density distribution can lead to a steep second power-law tail \citep{Schneider2022}. Conversely, angular momentum barriers or other mechanisms that slow gravitational collapse may result in a shallower second power-law \citep{Kritsuk2011}. While stellar feedback can lead to a second log-normal between the log-normal and power-law \citep{Tremblin2014,Das2017,Schneider2022}, a shallower second power-law at the highest column densities \citep{Tremblin2014,Schneider2022}, or over-density components \citep{Kainulainen2011,Schneider2012,Schneider2013,Ladjelate2020}. 

Simulations have shown that the parameters of the N-PDF can change over time. The log-normal widens as the initial density fluctuation enhances \citep{Ballesteros-Paredes2011,Ward2014,Schneider2015_1}. The transitional column density decreases towards certain values \citep{Ward2014,Burkhart2017}. The power-law part of the N-PDF emerges with the occurrence of star formation \citep{Ballesteros-Paredes2011,Kritsuk2011,Collins2012,Ward2014,Schneider2015_1,Burkhart2017}. The index of this power-law component decreases to values around 2 or 1.5 as star formation progresses \citep{Kritsuk2011,Ward2014,Girichidis2014,Burkhart2015} and is correlated with the efficiency of star formation \citep{Federrath2013}. Observations also see that the power-law index not only indicates the evolutionary stage of the molecular cloud \citep{Schneider2012,Russeil2013} but also varies with star formation efficiency \citep{Sadavoy2014,Kainulainen2014,Burkhart2018_863}, local young stellar object (YSO) counts \citep{Pokhrel2016}, and the fraction of Class 0 protostars \citep{Stutz2015}, suggesting that the power-law index can serve as an indicator of the circumstances and evolutionary state of molecular clouds. 

Located at a distance of $1.4\,\rm kpc$ \citep{Rygl2012}, Cygnus-X is one of the most active massive star formation regions in the Milky Way. Previous studies on Cygnus-X found it to have a total gas mass of $4.7\times 10^6 \, M_\odot$ \citep{Schneider2006}, and harbors numerous massive dense cores \citep{Motte2007,Cao2019}, OB associations \citep{Uyaniker2001}, and HII regions \citep{Wendker1991}. 
Previous studies have revealed that the North part of Cygnus-X exhibits typical log-normal + power-law N-PDF morphologies in dust continuum data, CO, and other molecular tracers \citep{Schneider2016}, and are also well fitted with double-log-normal + double-power-law models \citep{Schneider2022}.

In the first paper of our project, Surveys of Clumps, CorEs, and CoNdenSations in CygnUS-X (CENSUS), \citet{Cao2019} created a $5^\circ \times 6^\circ$ column density map of Cygnus-X. The map is generated from SED fittings to the dust continuum images from $Herschel$ \citep{Motte2010,Molinari2010,Hora2011}, and has a resolution of $18''.4$. Using the column density map from \citet{Cao2019}, we present a detailed analysis of the N-PDFs of Cygnus-X. With HI and YSO data, we obtain the N-PDF parameters and examine their relationship with star formation. We present our N-PDF results in Section \ref{sec:Me_Re}. The relation between N-PDF properties and star formation is analyzed in Section \ref{sec:analysis}. We discuss the importance of N-PDF in Section \ref{sec:discuss}. The main results are summarized in Section \ref{sec:sum}. 

\section{The PDFs of Cygnus-X}\label{sec:Me_Re}
\subsection{The N-PDF of Cygnus-X}\label{sec:npdfwhole}

\begin{figure}[htbp]
    \centering
    \includegraphics[width=\figwidth]{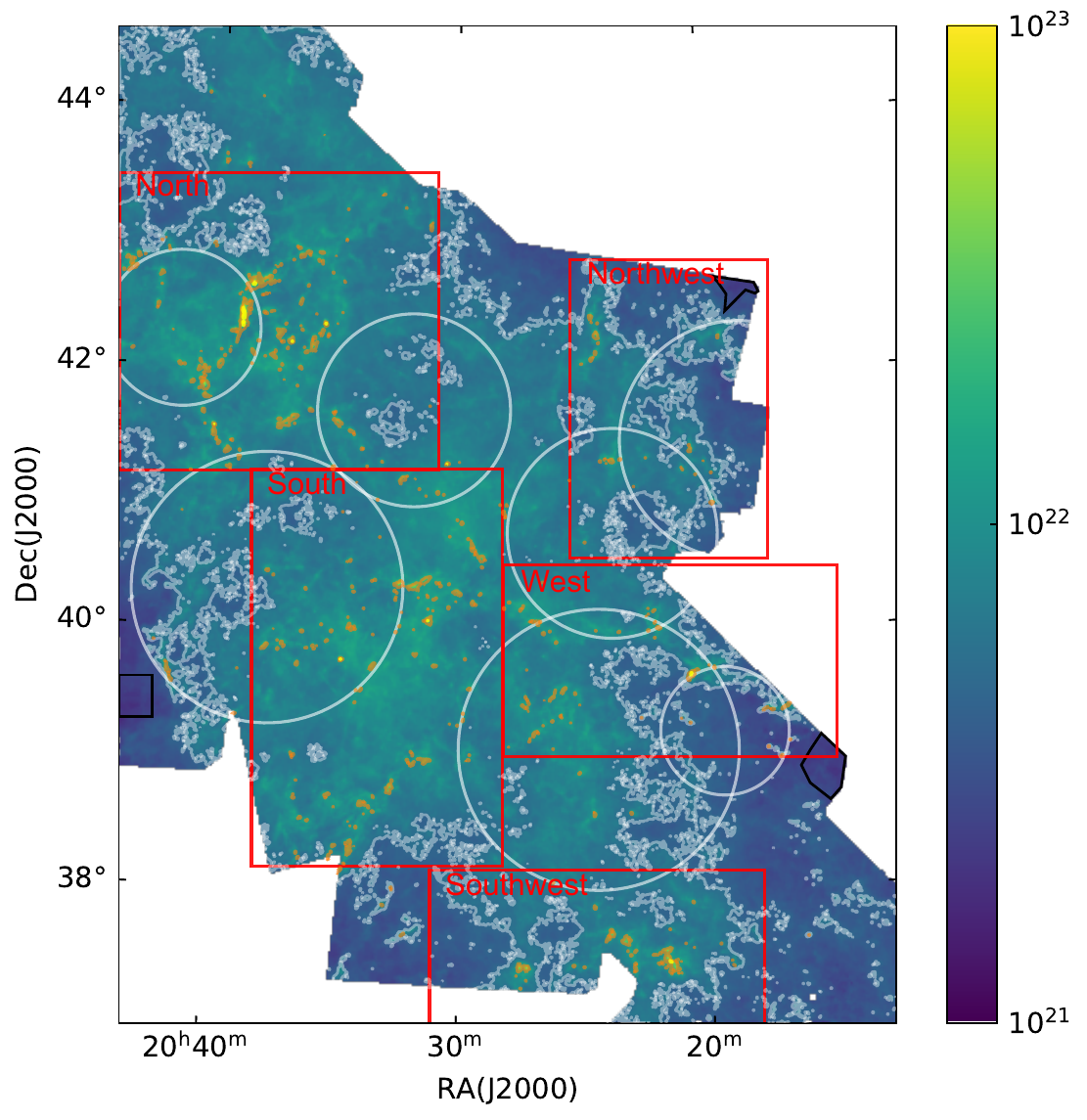}
    \caption{The column density map of Cygnus-X in [$\rm cm^{-2}$]. Red rectangles show the five sub-regions we selected. White contour outlines pixels with column density larger than $2.64\times10^{21}\,\rm{cm}^{-2}$, which are used in the N-PDF fitting. Black polygons show the three areas used for subtracting LOS contamination, their mean column density is $2.36\times 10^{21}\,\rm cm^{-2}$. White circles show the developed HII regions from \citet{Anderson2014}. The orange contour outlines structures with column density larger than $1.62\times10^{22}\,\rm{cm}^{-2}$, which are called dense structures in Section \ref{sec:TPclump}. The yellow contour outlines structures with column density larger than $1.23\times10^{23}\,\rm{cm}^{-2}$.}
    \label{fig:nmap}
\end{figure}

The Cygnus-X column density map we use is from \citet{Cao2019}. Applying SED fittings to the 160, 250, 350, and 500 $\rm{\mu m}$ mosaic images from Herschel \citep{Motte2010,Molinari2010,Hora2011}, they obtained the column density map with an angular resolution of $18''.4$. The resulting column density map has a scale of $5^\circ \times 6^\circ$, which is sufficiently large to encompass the relatively complete diffuse component of Cygnus-X. Most of the gas in Cygnus-X is contained within the white contour in Figure \ref{fig:nmap}, and the morphology of this contour is also relatively complete. Restricting our N-PDF analysis to this contour can avoid biases from incomplete boundaries \citep{Ossenkopf-Okada2016,Alves2017} and eliminates the lowest densities where subsequent line-of-sight (LOS) contamination subtraction introduces large uncertainties. To reduce the impact of LOS contamination, i.e., emission from dust tracing gas in the foreground/background \citep{Schneider2015_1,Ossenkopf-Okada2016}, we selected three regions that exhibit no discernible structure outside the white contour in Figure~\ref{fig:nmap} and subtracted their mean column density of $2.36 \times 10^{21}\,\rm cm^{-2}$ from the N-PDF. \footnote{In the following text, all column densities refer to the column densities after deducting the LOS contamination, unless specified.} The three regions are shown in black polygons in Figure \ref{fig:nmap}. Although the method of selecting the background is the same, the LOS contamination value we use is lower than that in \citet{Schneider2016,Schneider2022} by a factor of 2. They used a LOS contamination value of $5 \times 10^{21}\,\rm cm^{-2}$ for the Cygnus-X North and South regions, which is nearly a factor of 2 higher than ours. This difference is primarily due to the smaller field of view used in \citet{Schneider2016,Schneider2022}, which results in background regions selected with a higher density than those in our study. For Cygnus-X, which has a maximum column density reaching $10^{24}$ cm$^{-2}$, this LOS contamination value is relatively small. Therefore, the difference in LOS contamination values has a negligible impact on the overall shape and most parameters of the N-PDF, except for the most probable column density of the N-PDF $N_{\rm peak}$, which is below $10^{22}$ cm$^{-2}$. This minor difference does not affect the overall analysis of the N-PDF.

\begin{figure}[htbp]
    \centering
    \includegraphics[width=\figwidth]{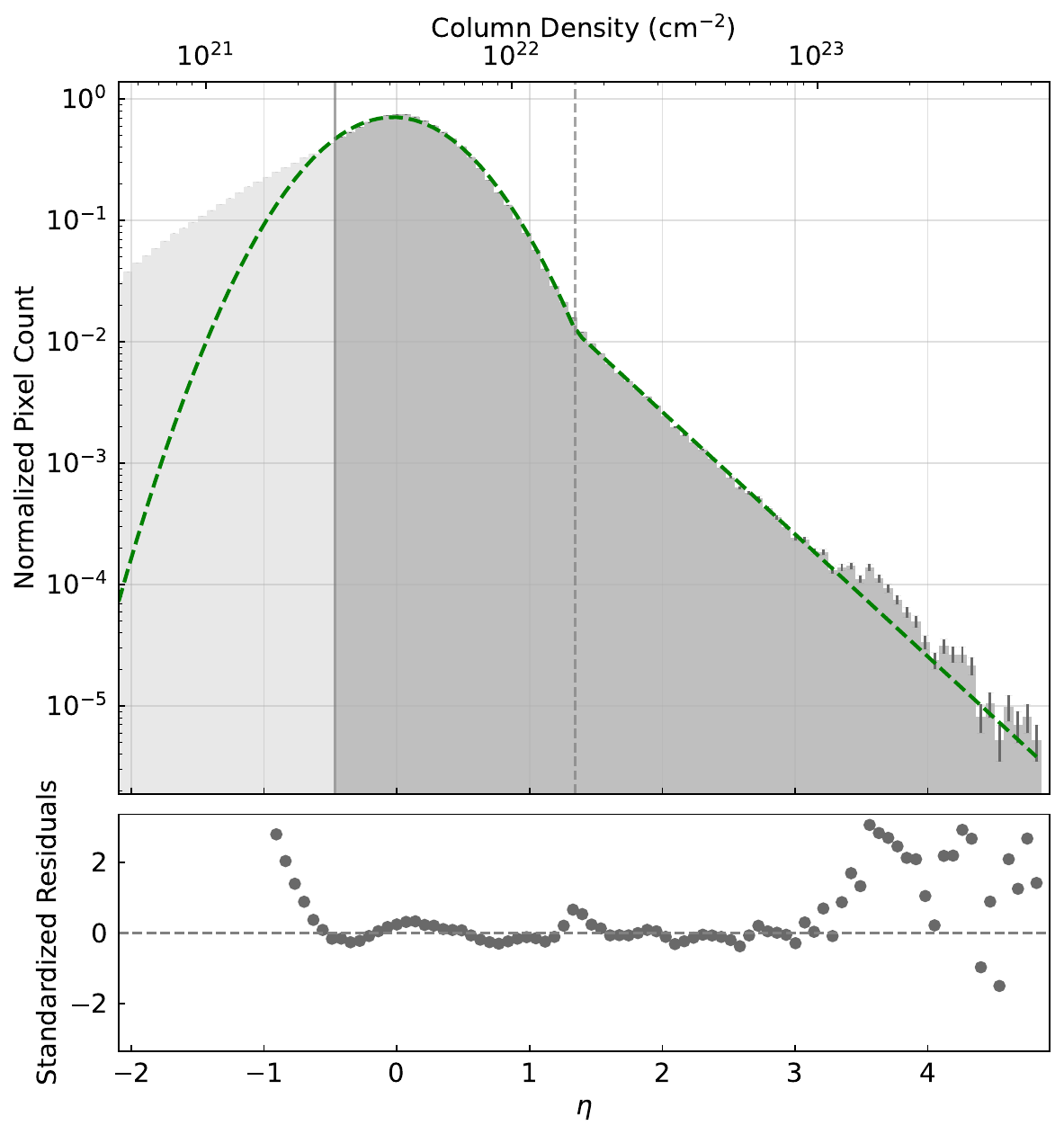}
    \caption{Histogram of the H$_2$ N-PDF of Cygnus-X. The solid vertical line marks $2.64\times10^{21}\,\mathrm{cm}^{-2}$, above which the N-PDF is fitted with a log-normal + power-law function. The green dashed line shows best-fit result to the N-PDF, the transition point between log-normal and power-law is shown in the dotted vertical line. The error bars on the histogram are derived from Poisson statistics and are small due to the large number of pixels used in the N-PDF. The standardized residuals of the fit, calculated in the logarithmic scale, are displayed in the lower panel. }
    \label{fig:npdf}
\end{figure}

We plot the N-PDF of Cygnus-X in Figure~\ref{fig:npdf}. The overall shape of the Cygnus-X N-PDF follows a lognormal plus power-law form. At the lowest column densities, an error power-law tail emerges due to the LOS contamination subtraction, consistent with the findings of \citet{Ossenkopf-Okada2016}. For column densities above the white contour at $2.64\times10^{21}\,\mathrm{cm}^{-2}$ (corresponding to $5.00\times10^{21}\,\mathrm{cm}^{-2}$ before LOS correction), we fit the N-PDF using the lognormal + power-law piecewise function given in Equation \ref{equ:piece}, 
\begin{equation}
    p(\eta)=
	\begin{cases}
	\frac{1}{\sqrt{2\pi}\sigma_{\eta}}{\rm C_{0}}\exp(-\frac{(\eta-N_{\rm peak})^{2}}{2\sigma_{\eta}^{2}})  &\eta\le\eta_{\rm TP} \\
	\exp({\rm C_1}\cdot \eta^{-s})  &\eta>\eta_{\rm TP}
	\end{cases}
	\label{equ:piece}
\end{equation}
where $\eta_{\rm TP}$ is the transitional column density, $\sigma_{\eta}$ is the standard deviation of the log-normal function. $s$ is the index of the power-law part. $\rm C_0$ and $\rm C_1$ are constants normalizing the piecewise function.

The fitting results of the N-PDF are listed in Table \ref{tab:wholeand5}. At low column densities, the N-PDF follows a log-normal distribution with $\sigma_{\eta}=0.48$ and a peak at $N_{\rm peak}=4.37\times 10^{21}$ $\rm cm^{-2}$. Above the transition point at $N_{\rm TP}=1.62\times 10^{22},\rm cm^{-2}$, the N-PDF exhibits a power-law tail with an index of $s=2.33$. These parameters show broad consistency with \citet{Schneider2022} despite differing methodologies: \citet{Schneider2022} analyzed the North and South regions separately using a double log-normal + double power-law model, while we study the entire Cygnus-X complex with a single transition. After accounting for differences in contamination subtraction, our $N_{\rm peak}$ aligns with their reported average peak extinction of 2.56 mag. Similarly, their characteristic $\sigma$ values (0.52 for both regions) closely match our $\sigma_{\eta}=0.48$. The transition points show minor discrepancies, their $N_{\rm TP,1}$ values (18.5 mag and 15.2 mag for North/South), likely due to their double-component fitting and differing contamination treatments.


An in Equation \ref{equ:forcing}, the log-normal part of the N-PDF is related to the forcing parameter $b$ and the sonic Mach number $M_s$ \citep{Burkhart2012}, 
\begin{equation}
\sigma_{\eta}=A\times \ln(1+b^2 M_s^2)
\label{equ:forcing}
\end{equation}
where $A$ is the scaling parameter, it has a best fit value of $0.11$ from simulated clouds obtained by \citet{Burkhart2012}. The sonic Mach number is defined as $M_s=\sqrt{3}\sigma_{v,\rm 1D}/c_s$. We adopt the velocity dispersion $\sigma_v$ of Cygnus-X as $\sigma_{v,\rm 1D}=4.99\times 10^3$ m/s from \citet{Spilker2022}. The sound speed $c_s$ can be calculated from the temperature, $c_s=\sqrt{\frac{k_{\rm b}T}{\mu m_{H}}}$, where we adopt the mean molecular mass $\mu=2.33$. Using the dust temperature map from \citet{Cao2019}, we obtain the temperature of the diffuse gas $T=17.65$ K, and get the sound speed $c_s=251.68$ m/s. From these values, we calculate the sonic Mach number $M_s=34.34$ and the forcing parameter $b=0.08$. 
The low value of $b$ suggests that the turbulence within Cygnus-X is predominantly driven by solenoidal forcing. The influence of magnetic fields, which has not been accounted for in our calculations, could be a contributing factor to the reduced $b$ value compared to the typical solenoidal value of $b$ = $\frac{1}{3}$. The effects of magnetic fields, as described by \citet{Molina2012}, are likely to adjust the value upwards, closer to the expected solenoidal range. Additionally, the choice of the scaling parameter $A = 0.11$ may also introduce a bias in our estimation of the low $b$ value for Cygnus-X. The scaling parameter is influenced by factors such as stellar feedback, the galactic environment, and the intrinsic properties of the cloud. Given the active and complex star-forming nature of Cygnus-X, it is possible that the actual $A$ value may deviate from the one derived from the simulated clouds in \citet{Burkhart2012}.

Above the log-normal part, the N-PDF has a clear transition at $N_{\rm TP}=1.62\times 10^{22}\,\rm cm^{-2}$ ($A_{\rm v}$= 17.2 mag). The transitional column density is higher than those typically observed in nearby molecular clouds, which range from $A_{\rm v} = 3-6$ mag \citep{Schneider2013,Schneider2015_1}. The $N_{\rm peak}$ of Cygnus-X at $4.37 \times 10^{21}$ cm$^{-2}$ further underscores its distinction from the nearby molecular clouds, which usually have peaks at $A_{\rm v}\le 1$ mag. This is consistent with the pattern identified across 29 molecular clouds by \citet{Schneider2022}, where clouds of massive star formation, including Cygnus-X North and South, have $N_{\rm TP}$ at $A_{\rm v}= 8-37$ mag and $N_{\rm peak}$ at $A_{\rm v}=3-3.8$ mag. These values are 7 and 3 times greater than those found in low-mass star-forming clouds. These characteristics reflect the inherent properties of Cygnus-X as a massive star formation region, potentially also related to its local environment. 

Above the transition, the N-PDF takes on a power-law shape with an index of $s=2.33$. Assuming spherical symmetry, for a source that has a power-law N-PDF with index $s$, we can calculate that it has a power-law density profile $\rho \propto r^{-\alpha}$ with $\alpha=1+2/s$ \citep[e.g.,][]{Kritsuk2011,Federrath2013}). Therefore, the power-law index of Cygnus-X indicates that the gases follow a density profile of $\rho\propto r^{-1.86}$ on average. It satisfies the self-gravitational collapse which has $\alpha=1.5-2$ in theories \citep{Larson1969,Penston1969,Shu1977}.

At $1.23\times 10^{23}\,\rm cm^{-2}$, the N-PDF shows a deviation from the power-law. As plotted in Figure \ref{fig:npdf}, the gas with column densities above $1.23\times 10^{23}\,\rm cm^{-2}$ distributes in the densest part in sub-regions North, West, and South (see Section \ref{sec:5reg} for the sub-regions). The N-PDFs of the sub-regions confirm this feature is prominent in the North and West regions. Such deviation from power-laws is commonly observed in molecular clouds and is sometimes seen as a second power-law \citep{Schneider2015, Schneider2016, Pokhrel2016}. It might be attributed to compression processes such as stellar feedback \citep{Tremblin2014,Schneider2022}.

\subsection{The HI N-PDF}\label{sec:HI}

\begin{figure}
    \centering
    \includegraphics[width=\figwidth]{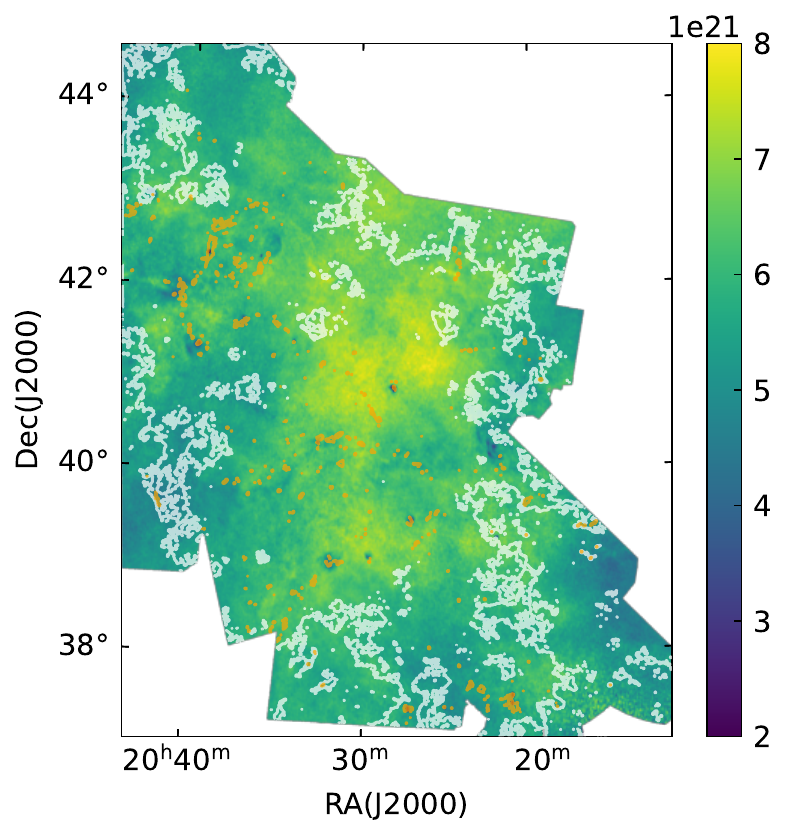}
    \caption{The HI column density map of Cygnus-X in $N_{\rm HI}$ [$\rm cm^{-2}$]. White and orange contours are from the dust continuum map and make the column densities of $2.64\times10^{21}\,\rm{cm}^{-2}$ and $1.62\times10^{22}\,\rm{cm}^{-2}$, respectively.}
    \label{fig:himap}
\end{figure}

We use the atomic hydrogen (HI) $21\,\mathrm{cm}$ spectral line data from the Canadian Galactic Plane Survey \citep[CGPS,][]{Taylor2003} to obtain the HI N-PDF of Cygnus-X. The data have a spatial resolution of $88\farcs8$ at $+40\fdg8$ declination, and a velocity resolution of $0.82\,\mathrm{km\,s^{-1}}$. 
While the molecular component of Cygnus-X, traced by CO emission, spans a local standard of rest (LSR) velocity range of $-10$ to $+20\,\mathrm{km\,s^{-1}}$ \citep{Schneider2006}, the atomic hydrogen reveals a more extended kinematic structure. Based on the CGPS data, the HI emission associated with Cygnus-X spans a broad velocity range of $-20$ to $+30\,\mathrm{km\,s^{-1}}$. We integrate over this range to capture the complete atomic envelope associated with Cygnus-X. 
We observed HI self-absorption (HISA) features in the HI data, primarily at $N_{\mathrm{HI}}\le 4\times10^{21}\,\mathrm{cm^{-2}}$. These regions occupy only a small fraction of the area, so we estimate the HI column density using the optically thin approximation as given in Equation~\ref{equ:1}.
\begin{equation}
	\label{equ:1}
    N({\rm HI}) \, ({\rm cm^{-2}})=1.823\times 10^{18}\int T_{\rm B}(v)dv \, (\rm K \, km/s)
\end{equation}
The obtained HI column density map is shown in Figure \ref{fig:himap}. All emissions above three times the theoretical RMS noise, i.e., above $6.87$ K, are used in the process. 

\begin{figure}
    \centering
    \includegraphics[width=\figwidth]{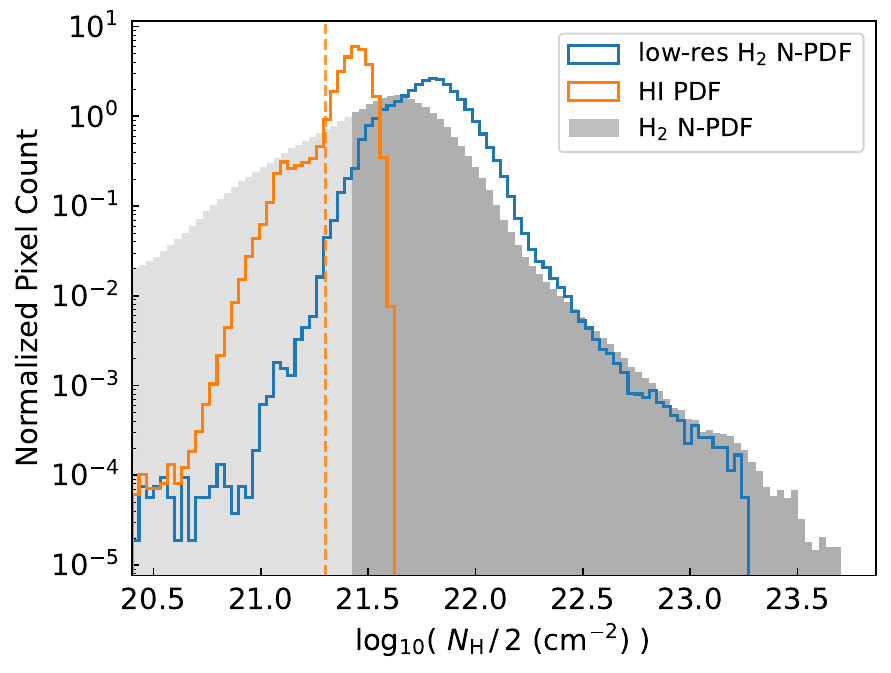}
    \caption{HI and H$_2$ N-PDFs of Cygnus-X. The orange line shows the HI N-PDF. The grey region represents the H$_2$ N-PDF analyzed throughout this work. The blue line shows the same H$_2$ N-PDF after convolution to the HI angular resolution and with no LOS contamination correction applied. The vertical orange line marks $N\rm_H = 4\times10^{21}\, cm^{-2}$, below which the HI data are visibly affected by HISA. To make fair comparison of the N-PDFs, the $x$ axis is in $N_{\rm H}/2$. }
    \label{fig:hipdf}
\end{figure}

We compare the Cygnus‑X HI N-PDF and the H$_2$ N‑PDF in Figure~\ref{fig:hipdf}. For clarity, in addition to the H$_2$ N‑PDF as in Figure~\ref{fig:npdf} (in grey), we also convolved the dust column density map to the same spatial resolution and pixel scale as the HI data, without subtracting the line‑of‑sight contamination (in blue). The HI N-PDF exhibits a dip below $N_{\rm HI}\sim4\times10^{21}\,\mathrm{cm^{-2}}$, which is possibly related to the underestimation of the HI column density due to HISA. Ignoring this feature, the overall shape of the HI N-PDF resembles a log‑normal distribution but is skewed toward the high column density end, and it cannot be adequately fitted by either a log‑normal or a normal function. The peak of the HI N-PDF occurs at $N_{\rm HI}=5.5\times10^{21}\,\mathrm{cm^{-2}}$, with most of the gas confined to a narrow column density range between $N_{\rm HI}=4\times10^{21}$ and $8\times10^{21}\,\mathrm{cm^{-2}}$. Both the peak and the truncation at the high‑density end lie below the peak of the H$_2$ N‑PDF, suggesting a link between the HI N-PDF morphology and the HI‑to‑H$_2$ transition. Note that the column densities presented here are LOS integrated values, and therefore differ conceptually from the theoretical HI‑to‑H$_2$ transition thresholds derived in analytic models \citep[e.g.,][]{Krumholz2008a}. The width of the HI N-PDF is significantly narrower than that of the H$_2$ N‑PDF, it is also observed in nearby molecular clouds such as Perseus and Orion \citep{Burkhart2015b,Imara2016}.

\subsection{The five sub-regions}\label{sec:5reg}

We divide Cygnus-X into five sub-regions (Figure~\ref{fig:nmap}) based on distinct morphological and star-forming properties identified in previous studies \citep{Schneider2006,Motte2007}. North (N) is an active massive star-forming region dominated by dense filamentary structures, hosting prominent sites like DR21 and W75N. South (S) is located south of the Cyg OB2 association. This region is clumpy, less evolved, and exhibits weaker star formation activity \citep{Takekoshi2019}. West (W) is mainly a filamentary extension connected to the S region, and is often regarded as part of ``Cygnus-X South'' \citep[e.g.,][]{Schneider2006}. Cyg OB9 and dark cloud L889 are in this region. Northwest (NW) is in proximity to Cyg OB8, features relatively diffuse and filamentary gas. Its most prominent feature is the dark cloud L897. Southwest (SW) is at west of the southernmost end of Cygnus-X, far away from other regions, and the gas is concentrated in several massive star-forming clumps with high densities, such as S106.

\begin{figure*}
\centering
\gridline{\fig{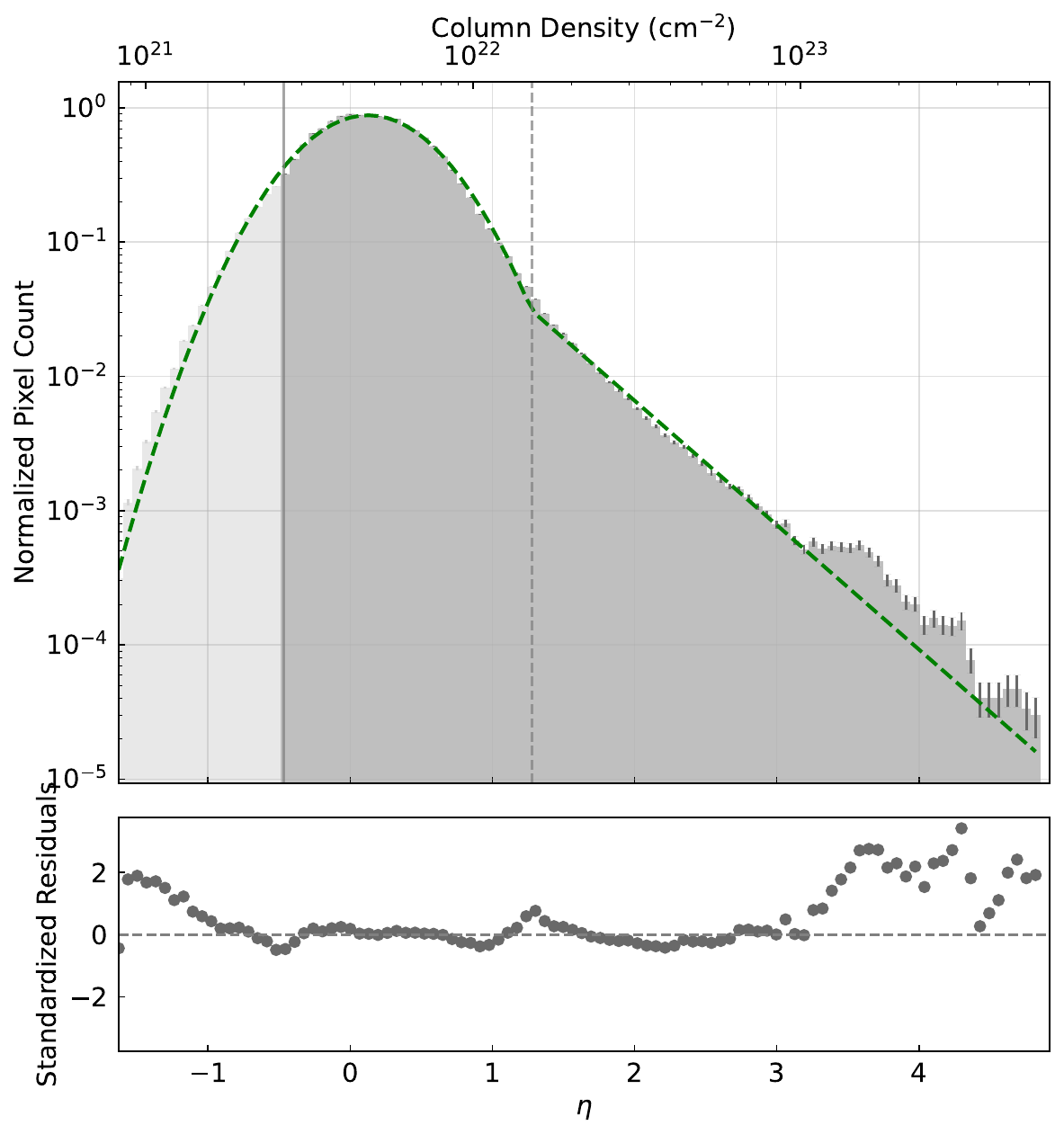}{0.3\textwidth}{North}
          \fig{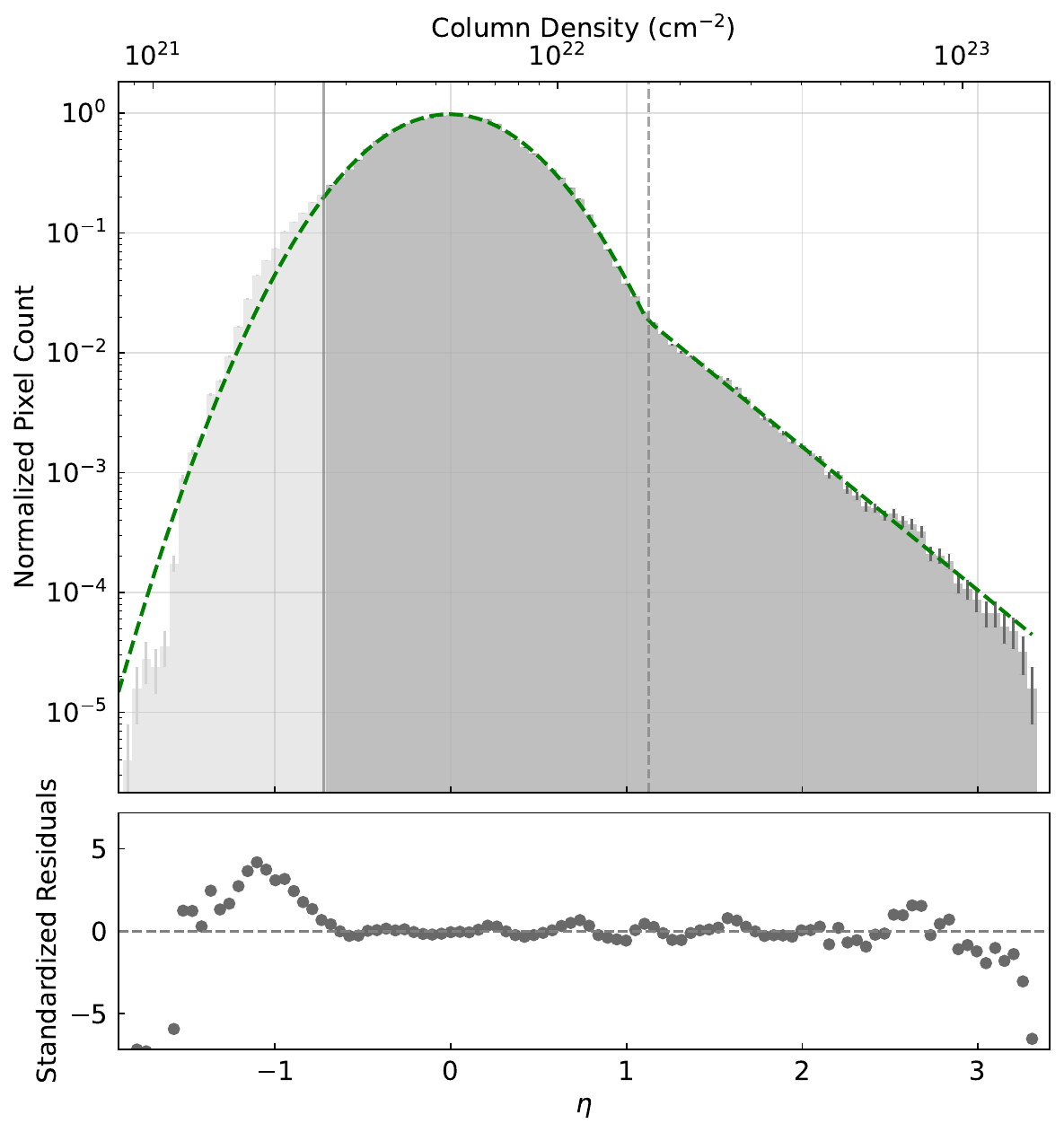}{0.3\textwidth}{South}
          \fig{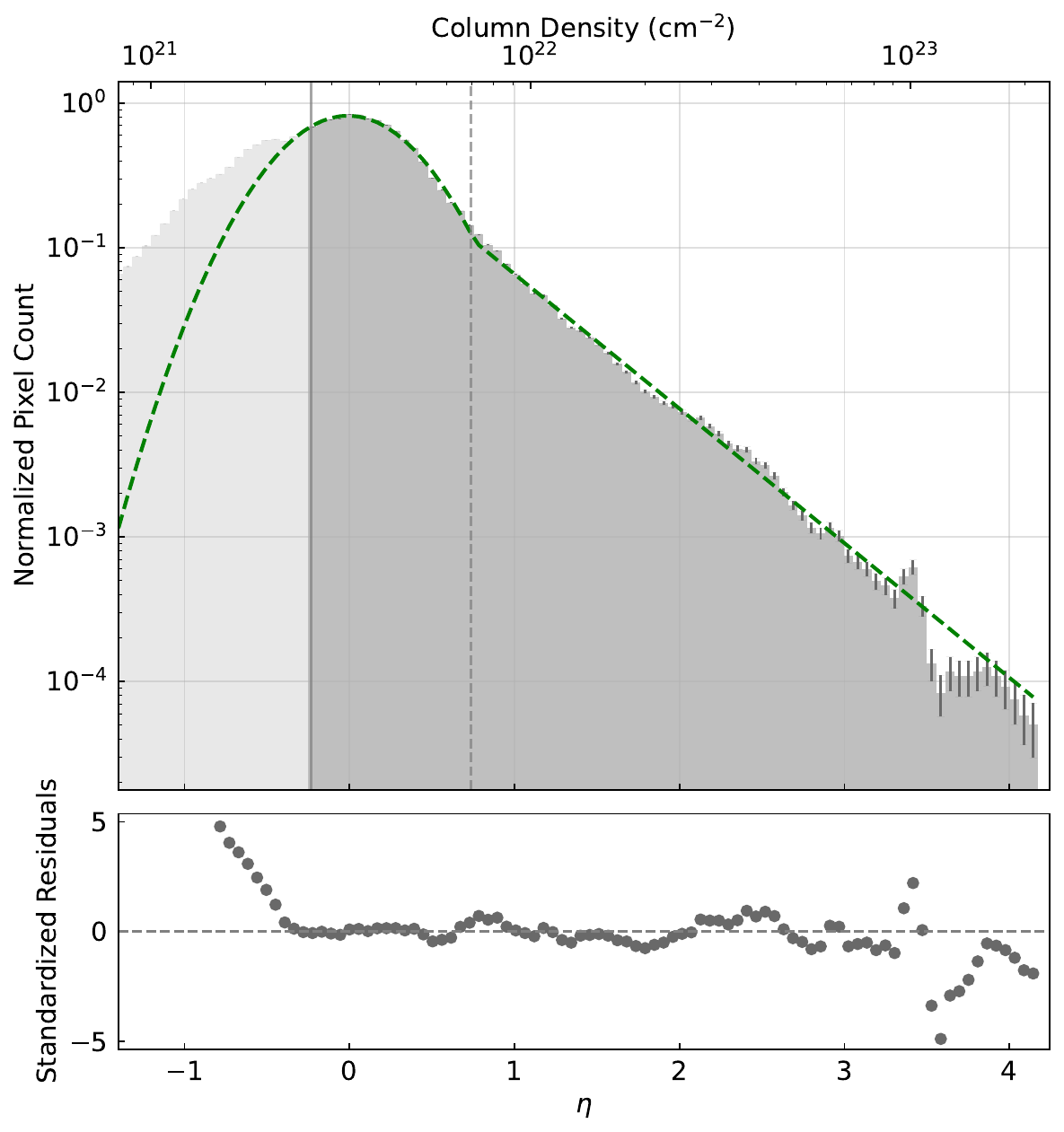}{0.3\textwidth}{Southwest}}
\gridline{\fig{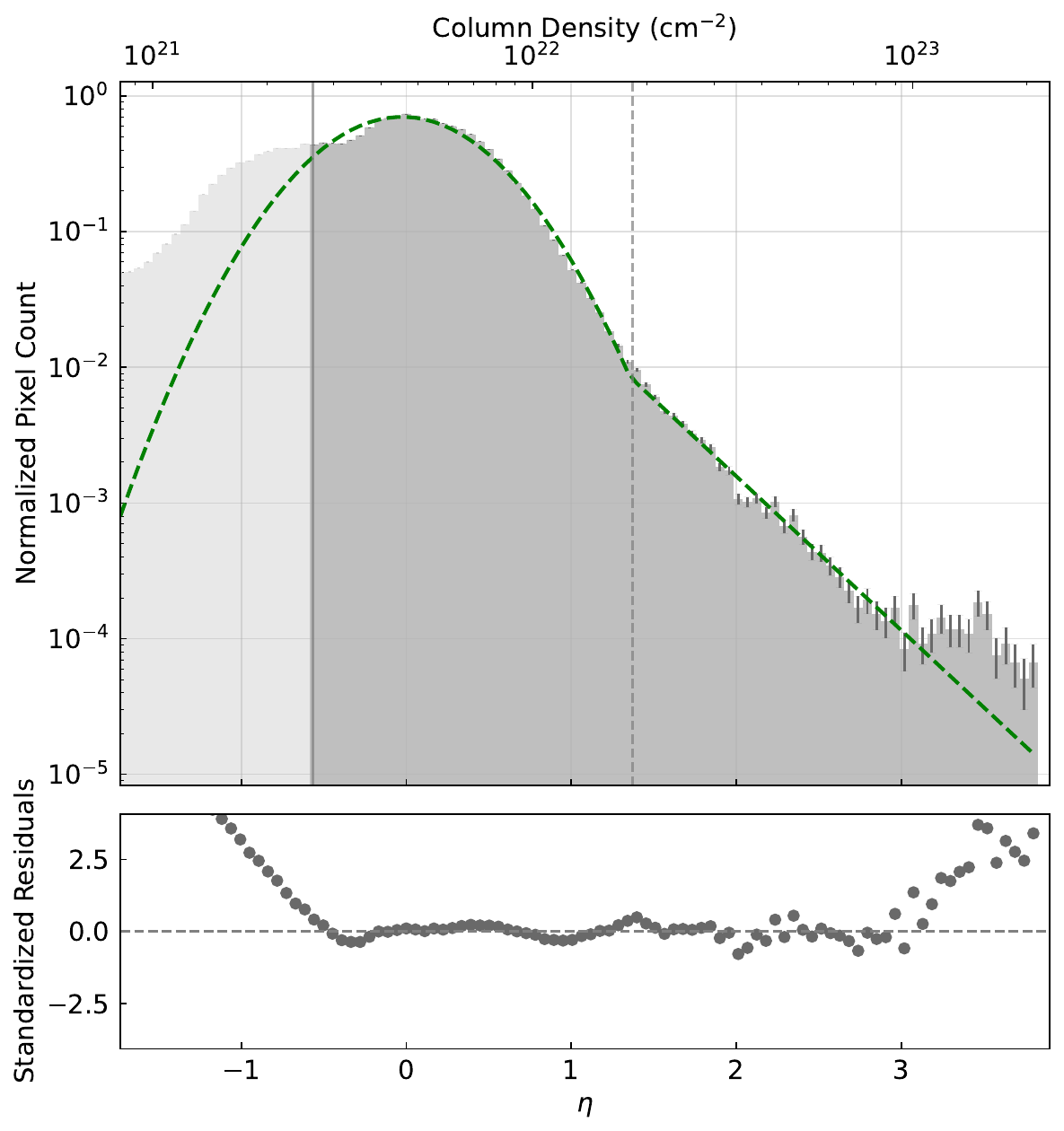}{0.3\textwidth}{West}
          \fig{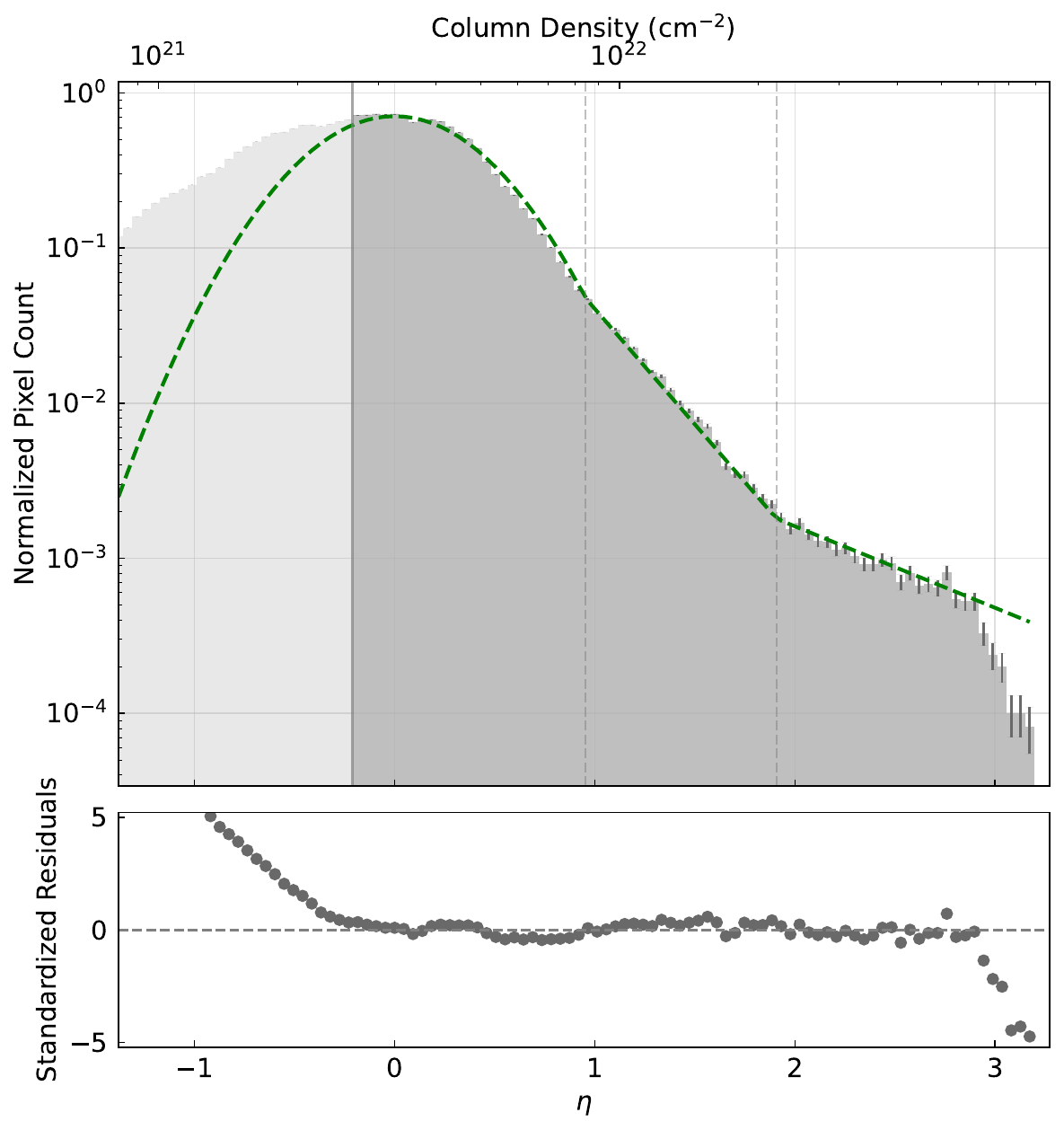}{0.3\textwidth}{Northwest}}
\caption{Similar to Figure~\ref{fig:npdf} but for N-PDFs of the five sub-regions in Cygnus-X. For the first four panels, the green curves show log-normal + power-law fitting to the N-PDFs, and the vertical dashed lines mark the transition points. While for region ``Northwest'', the N-PDF is fitted with a log-normal + double power-law function, with vertical dashed lines showing the two transition points between the three parts. }
\label{fig:npdf_5}
\end{figure*}

Four out of the five sub-regions have N-PDFs well described by log-normal + power-law shapes. We fit them with log-normal + power-law functions. The NW region displays an additional excess between the log-normal and the power-law components. We fit its N-PDF with log-normal + double power-laws. All the results are shown in Figure~\ref{fig:npdf_5} and Table~\ref{tab:wholeand5}. 
Across all sub-regions, the log-normal widths lie in a narrow range $\sigma_\eta = 0.39$–$0.46$. Regions with larger $\sigma_\eta$ (N and W) are suggested to have stronger turbulence than S, NW, and SW. The power-law indices of the four sub-regions with single power-law tails span $s=2.14$–$2.77$, corresponding to volume-density profiles $\rho\propto R^{-\alpha}$ with $\alpha = 2/(s-1)$ ranging from 1.72 to 1.93.

The power-law index of NW at the high-density end is 1.21 ($\alpha=2.65$), indicating a centrally concentrated mass distribution steeper than pure free-fall collapse. Such a steep profile points to additional external compression acting on the gas \citep{Kainulainen2009,Schneider2015}. Another power-law component is fitted at intermediate column densities ($8.5\times10^{21}\,{\rm cm^{-2}}\lesssim N_{\rm H_2}\lesssim2.2\times10^{22}\,{\rm cm^{-2}}$) with slope $s=3.40$. Its origin can be interpreted in two ways: (i) a second gravity-dominated component whose characteristic density and index differ from the high-density one, leading to two distinct power-laws due to superposition in the PDF \citep{Chen2018}; or (ii) a pressure-driven over-density component corresponding to the ``ridge'' or ``shell'' compressed by the expanding shell of the HII region. \citep{Schneider2012,Schneider2013,Tremblin2014,Das2017,Ladjelate2020}. Due to the presence of two extended HII regions around NW \citep[white circles in Figure~\ref{fig:nmap}, data from][]{Anderson2014} and that the four dense regions in NW (region 12-15 in Section~\ref{sec:32reg}) all have power-law indices less than 2, we favour the second scenario. The intermediate-density power-law could be a pressure-driven over-density component compressed by the ionization fronts and stellar winds from the neighboring HII regions. Together, these factors indicate that NW is subject to strong stellar feedback, which is imprinted in its distinct N‑PDF.

Three out of four regions with log-normal + power-law PDFs have $N_{\rm TP}$ around $1.6\times 10^{22}$ cm$^{-2}$, similar to that of the whole Cygnus-X. The Southwest region has a lower $N_{\rm TP}$, the $\sigma_\eta$ and $N_{\rm H,tran}$ of the region are also the lowest. This may be due to its greater distance from the Cygnus-X center, resulting in slightly different environmental conditions compared to the other sub-regions. The low $N_{\rm TP}$ of Southwest also results in a high dense-gas mass fraction $f_{\rm dense}$, which is defined as the fraction of mass with column density above the transitional point. 

\begin{deluxetable*}{lcccccc}
\tablecaption{Properties of the Cygnus-X and the five sub-regions.}
\tablewidth{0pt}
\tablehead{
\colhead{Region} & \colhead{Center(RA;Dec)} & \colhead{\(\sigma_{\eta}\)} & \colhead{\(N_{\rm peak}\)} &  \colhead{\(s\)} & \colhead{\(N_{\rm TP}\)} & \colhead{\(f_{\rm dense}\)}
\\
\colhead{} & \colhead{($^h$:$^m$:$^s$; $^\circ$:$'$:$''$)} & \colhead{} & \colhead{($\rm cm^{-2}$)} & \colhead{} & \colhead{($\rm cm^{-2}$)} & \colhead{(\%)} 
}
\decimalcolnumbers
\startdata
Cygnus-X & - & 0.48 & 4.37$\times 10^{21}$ & 2.33 & $1.62\times 10^{22}$ & 3.77\\
North & 20:37:32.380; +42:20:38.91 & 0.45 & 4.15$\times 10^{21}$ & 2.14 & $1.57\times 10^{22}$ & 7.61\\
South & 20:33:14.890; +39:41:14.77 & 0.40 & 5.37$\times 10^{21}$ & 2.77 & $1.65\times 10^{22}$ & 2.86\\
West & 20:21:27.993; +39:44:00.80 & 0.46 & 4.78$\times 10^{21}$ & 2.64 & $1.71\times 10^{22}$ & 2.71\\
Northwest & 20:21:19.644; +41:40:57.92 & 0.41 & 3.29$\times 10^{21}$ & 3.40, 1.21 & $8.54\times 10^{21}$, $2.22\times 10^{22}$ & - \\
Southwest & 20:24:31.245; +37:30:59.99 & 0.39 & 3.36$\times 10^{21}$ & 2.14 & $7.22\times 10^{21}$ & 22.89\\
\enddata
\label{tab:wholeand5}
\tablecomments{Columns 3-6 give the standard deviation of the log-normal, the most probable column density, the power-law index, and the transitional column density of the N-PDF, respectively. The dense-gas mass fraction is shown in Columns 6. Region Northwest is fitted with log-normal + double power-law, yielding two sets of $s$ and $N_{\rm TP})$. Because the physical meaning of the intermediate-density power-law in Northwest is uncertain, no $f_{\rm dense}$ is provided for this region.}
\end{deluxetable*}

\subsection{The 32 dense regions}\label{sec:32reg}

\begin{figure}
    \centering
    \includegraphics[width=\figwidth]{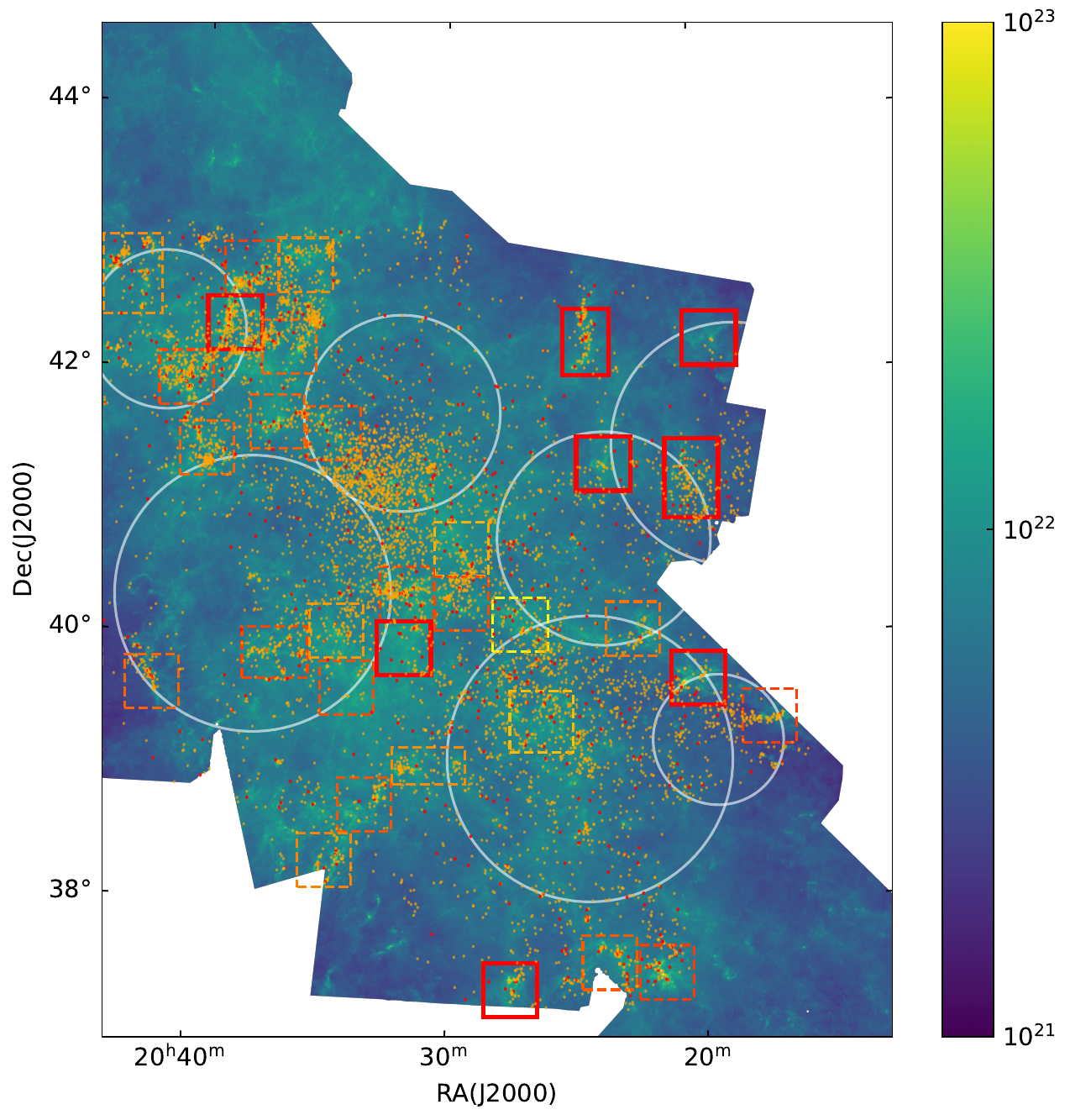}
    \caption{Same as Figure \ref{fig:nmap} but with different legends. The dashed and solid rectangles represent the 32 dense regions selected by $N>1.62\times 10^{22}\,\rm cm^{-2}$, colors of the rectangles from red to yellow represent power-law indices of the regions' N-PDFs from shallow to steep. Dense regions with power-law index $s<2$ are marked with solid red rectangles. The Class-I YSOs are shown in red dots, while the other YSOs are shown in orange dots. White circles show the developed HII regions in Cygnus-X. }
    \label{fig:nmap_32}
\end{figure}

\begin{figure*}
    \centering
    \includegraphics[width=0.7\linewidth]{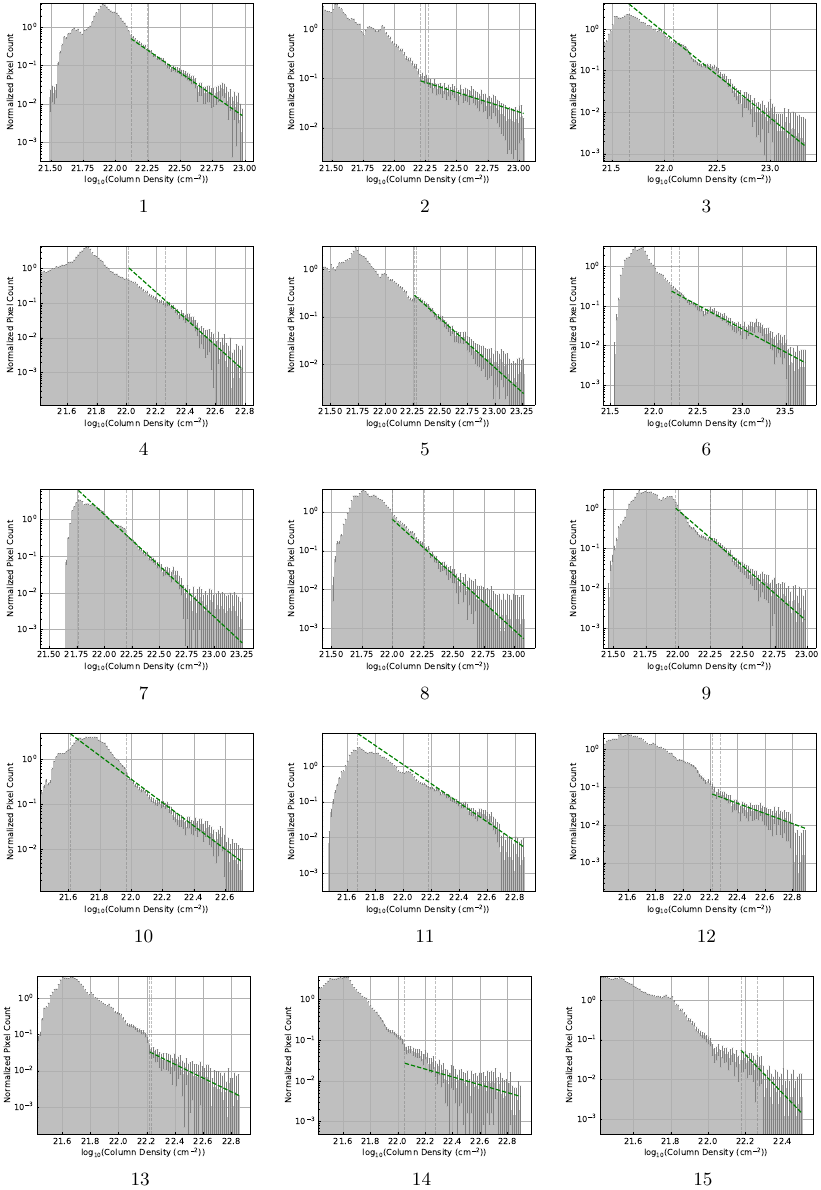}
    \caption{N-PDFs of the 32 dense regions in Cygnus-X. The green dotted lines show power-law fittings toward the high column density part of the N-PDFs, with vertical dashed lines showing the $1$ and $3\sigma$ confidence intervals of the fittings.}
    \label{fig:npdf32}
\end{figure*}

We also select the 32 dense regions in Figure \ref{fig:nmap_32}. These regions are rectangles with radii of about $10\,\rm pc$ ($0.41\,\rm deg$), covering most areas with column density larger than $1.62\times 10^{22}\,\rm cm^{-2}$. 
The low column density parts of the 32 dense regions' N-PDFs cannot be fitted by log-normal due to contour incompleteness \citep{Ossenkopf-Okada2016, Alves2017}. We perform power-law fittings on the high column density end of these N-PDFs, and the results are in Figure~\ref{fig:npdf32} and Table~\ref{tab:table32}. 

\begin{figure*}
    \centering
    \includegraphics[width=0.7\linewidth]{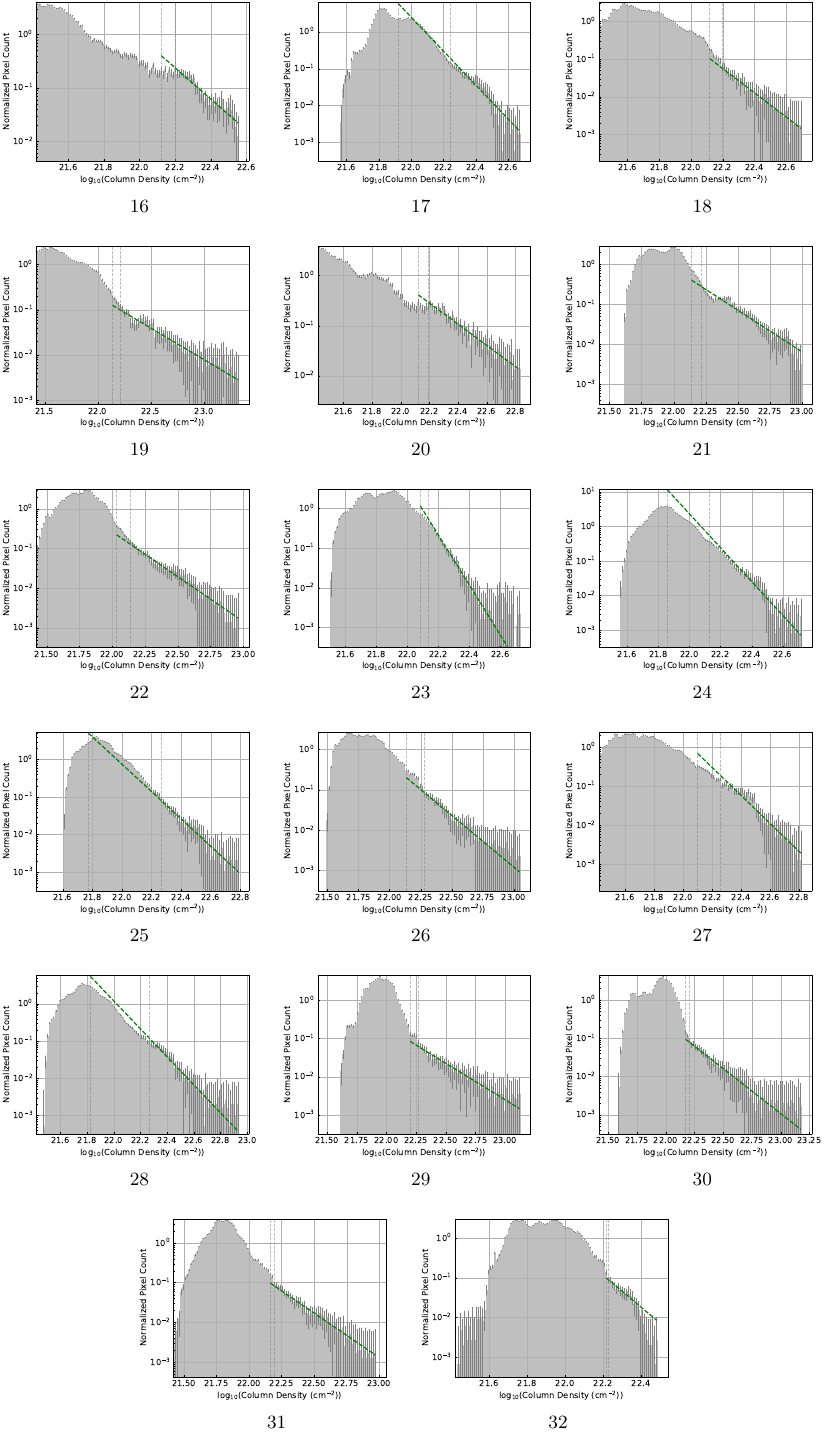}
    \caption{Same as Figure \ref{fig:npdf32}. }
    \label{fig:npdf32_1}
\end{figure*}

\begin{deluxetable*}{lccccccccc}
\tablecaption{Properties of the 32 dense regions.}
\label{tab:table32}
\tablewidth{0pt}
\tablehead{
\colhead{Region} & \colhead{Center(RA;Dec)} & \colhead{Size} & \colhead{$M_{\rm reg}$} & \colhead{\(s\)} & \colhead{\(N_{\rm TP}\)} & \colhead{\(f_{\rm dense}\)} & \colhead{\(N_{\rm max}\)} & \colhead{\(\Sigma_{\rm YSO}\)} & \colhead{\(f_{\rm ClassI YSO}\)} \\
\colhead{} & \colhead{($^h$:$^m$:$^s$; $^\circ$:$'$:$''$)} & \colhead{($\rm pc^2$)} & \colhead{($M_{\odot}$)} & \colhead{} & \colhead{($\rm cm^{-2}$)} & \colhead{(\%)} & \colhead{($\rm cm^{-2}$)} & \colhead{(/\(100\rm pc^2\))} & \colhead{(\%)}
}
\decimalcolnumbers
\startdata
1 & 20:40:40.701; +41:55:00.61 & 100.0 & $1.51\times 10^{4}$ & 2.33 & $1.75\times 10^{22}$ & 14.5 & $9.69\times 10^{22}$ & 254.0 & 8.7 \\
2 & 20:27:30.842; +37:18:32.92 & 100.0 & $6.6\times 10^{3}$ & 0.8 & $1.88\times 10^{22}$ & 19.5 & $1.11\times 10^{23}$ & 47.0 & 14.9 \\
3 & 20:21:31.077; +37:25:55.65 & 100.0 & $1.2\times 10^{4}$ & 2.05 & $1.21\times 10^{22}$ & 34.7 & $2.17\times 10^{23}$ & 64.0 & 25.0 \\
4 & 20:43:02.036; +42:41:03.31 & 160.6 & $1.51\times 10^{4}$ & 3.82 & $1.83\times 10^{22}$ & 5.3 & $6.13\times 10^{22}$ & 103.4 & 11.4 \\
5 & 20:38:07.844; +42:45:26.55 & 100.0 & $1.23\times 10^{4}$ & 2.07 & $1.92\times 10^{22}$ & 24.4 & $1.87\times 10^{23}$ & 140.0 & 12.9 \\
6 & 20:38:45.520; +42:20:29.41 & 100.0 & $1.82\times 10^{4}$ & 1.19 & $1.94\times 10^{22}$ & 37.7 & $5.4\times 10^{23}$ & 232.0 & 7.8 \\
7 & 20:36:30.432; +42:10:08.16 & 100.0 & $1.42\times 10^{4}$ & 2.8 & $1.6\times 10^{22}$ & 16.2 & $1.82\times 10^{23}$ & 169.0 & 5.9 \\
8 & 20:39:44.963; +41:23:00.91 & 100.0 & $1.18\times 10^{4}$ & 2.84 & $1.83\times 10^{22}$ & 6.5 & $1.22\times 10^{23}$ & 198.0 & 3.0 \\
9 & 20:36:54.675; +41:35:58.11 & 100.0 & $1.19\times 10^{4}$ & 2.79 & $1.76\times 10^{22}$ & 10.3 & $9.81\times 10^{22}$ & 75.0 & 12.0 \\
10 & 20:34:37.375; +41:31:10.92 & 100.0 & $9.47\times 10^{3}$ & 2.59 & $9.38\times 10^{21}$ & 15.1 & $5.19\times 10^{22}$ & 139.0 & 6.5 \\
11 & 20:23:41.917; +37:30:35.02 & 100.0 & $1.01\times 10^{4}$ & 2.67 & $1.52\times 10^{22}$ & 16.4 & $7.38\times 10^{22}$ & 78.0 & 19.2 \\
12 & 20:24:23.459; +42:13:06.52 & 104.5 & $8.9\times 10^{3}$ & 1.34 & $1.87\times 10^{22}$ & 7.9 & $7.93\times 10^{22}$ & 96.7 & 9.9 \\
13 & 20:23:43.352; +41:17:36.45 & 100.0 & $8.53\times 10^{3}$ & 1.9 & $1.72\times 10^{22}$ & 3.2 & $7.23\times 10^{22}$ & 32.0 & 9.4 \\
14 & 20:20:10.605; +41:10:39.80 & 146.6 & $8.76\times 10^{3}$ & 0.96 & $1.87\times 10^{22}$ & 4.2 & $7.94\times 10^{22}$ & 104.4 & 5.9 \\
15 & 20:19:19.856; +42:14:03.79 & 100.0 & $5.05\times 10^{3}$ & 1.95 & 
$1.11\times 10^{22}$ & 3.3 & $3.2\times 10^{22}$ & 9.0 & 0.0 \\
16 & 20:41:37.544; +39:36:08.08 & 100.0 & $4.05\times 10^{3}$ & 2.92 & $1.59\times 10^{22}$ & 13.1 & $3.61\times 10^{22}$ & 60.0 & 13.3 \\
17 & 20:29:26.587; +40:39:00.49 & 100.0 & $1.35\times 10^{4}$ & 4.59 & $1.76\times 10^{22}$ & 4.2 & $4.76\times 10^{22}$ & 107.0 & 7.5 \\
18 & 20:22:36.995; +40:02:25.27 & 100.0 & $8.56\times 10^{3}$ & 3.21 & $1.56\times 10^{22}$ & 3.0 & $4.99\times 10^{22}$ & 56.0 & 3.6 \\
19 & 20:20:04.490; +39:39:35.77 & 100.0 & $7.48\times 10^{3}$ & 1.39 & $1.64\times 10^{22}$ & 18.3 & $2.14\times 10^{23}$ & 66.0 & 9.1 \\
20 & 20:17:17.659; +39:21:26.39 & 100.0 & $2.91\times 10^{3}$ & 2.12 & $1.56\times 10^{22}$ & 23.5 & $6.83\times 10^{22}$ & 74.0 & 2.7 \\
21 & 20:31:36.035; +40:18:31.58 & 100.0 & $1.54\times 10^{4}$ & 2.09 & $1.65\times 10^{22}$ & 15.4 & $9.92\times 10^{22}$ & 201.0 & 7.5 \\
22 & 20:29:26.052; +40:14:14.72 & 100.0 & $1.04\times 10^{4}$ & 2.26 & $1.38\times 10^{22}$ & 8.1 & $9.26\times 10^{22}$ & 135.0 & 5.2 \\
23 & 20:27:04.548; +40:04:31.62 & 102.6 & $1.23\times 10^{4}$ & 6.29 & $1.36\times 10^{22}$ & 7.2 & $5.36\times 10^{22}$ & 111.1 & 13.2 \\
24 & 20:26:16.099; +39:20:21.49 & 133.2 & $1.64\times 10^{4}$ & 4.9 & $1.35\times 10^{22}$ & 8.3 & $5.27\times 10^{22}$ & 116.4 & 9.7 \\
25 & 20:35:55.203; +42:47:28.07 & 100.0 & $1.21\times 10^{4}$ & 3.65 & $1.84\times 10^{22}$ & 2.9 & $6.19\times 10^{22}$ & 137.0 & 1.5 \\
26 & 20:33:11.389; +38:42:24.69 & 100.0 & $1.13\times 10^{4}$ & 2.56 & $1.9\times 10^{22}$ & 6.3 & $1.13\times 10^{23}$ & 63.0 & 4.8 \\
27 & 20:34:41.953; +38:16:49.55 & 100.0 & $8.44\times 10^{3}$ & 3.58 & $1.8\times 10^{22}$ & 8.7 & $6.61\times 10^{22}$ & 69.0 & 11.6 \\
28 & 20:30:41.067; +39:00:06.62 & 92.8 & $1.03\times 10^{4}$ & 3.76 & $1.84\times 10^{22}$ & 5.2 & $8.68\times 10^{22}$ & 92.7 & 1.2 \\
29 & 20:31:41.711; +39:53:37.55 & 100.0 & $1.5\times 10^{4}$ & 1.91 & $1.85\times 10^{22}$ & 4.9 & $1.35\times 10^{23}$ & 65.0 & 15.4 \\
30 & 20:33:56.345; +39:35:23.96 & 100.0 & $1.42\times 10^{4}$ & 2.35 & $1.61\times 10^{22}$ & 4.5 & $1.52\times 10^{23}$ & 26.0 & 11.5 \\
31 & 20:36:48.420; +39:50:56.62 & 114.1 & $1.27\times 10^{4}$ & 2.24 & $1.56\times 10^{22}$ & 5.7 & $9.49\times 10^{22}$ & 113.9 & 6.2 \\
32 & 20:34:22.852; +40:01:09.54 & 100.0 & $1.29\times 10^{4}$ & 4.04 & $1.68\times 10^{22}$ & 1.9 & $3.07\times 10^{22}$ & 96.0 & 5.2 \\
\enddata
\tablecomments{Column 3-4 gives the total size and mass ($M_{\rm reg}$) of the dense region, obtained by summing values over the entire rectangular area. Columns 5-8 give the power-law index, the transitional column density, the dense-gas mass fraction, and the maximum column density, respectively. The dense-gas mass fraction is defined as the ratio between the mass of gas above the local transitional column density and the total mass. The YSO surface density (i.e., the normalized YSO count), and the Class-I YSO fraction are shown in Columns 9 and 10.}
\end{deluxetable*}

All of the 32 dense regions own power-law N-PDFs at high column density
. The power-law indices range from 0.80 to 6.29, with a mean of 2.78. Assuming spherical symmetry, these indices correspond to $\rho\propto R^{-\alpha}$ profiles with $\alpha=1.32-3.50$, with a mean of 1.72. 
In Figure \ref{fig:nmap_32}, the 32 dense regions are color-coded by their power-law indices $s$ from red (low $s$) to yellow (high $s$). Due to the intricate shape of Cygnus-X, the figure does not exhibit any clear spatial pattern of $s$ variation. Eight dense regions with $s<2$ are highlighted in solid red rectangles in Figure \ref{fig:nmap_32}. These regions have power-law index $0.80\le s\le1.95$ (corresponding to steep density profiles with $2.03\le\alpha\le 3.50$), implying a highly concentrated mass distribution. It may suggest that they are undergoing strong external compression, potentially from stellar feedback. Similarly, due to the intricate shape of Cygnus-X, we do not see the eight regions have distinctive spatial positions, except for that all the four dense regions located within NW have $s<2$. 

The dashed vertical lines in Figure \ref{fig:npdf32} show the column densities where the N-PDFs deviate from the 1 and 3$\sigma$ confidence intervals of the power-law fittings. Through visual judgment, we find the deviations from $1\sigma$ confidence intervals describe the transition better, thus, we consider them as the transitional column densities of the 32 dense regions. These column densities vary from $(0.94-1.94)\times 10^{22}\,\rm cm^{-2}$, and the mean is at $1.67\times 10^{22}\,\rm cm^{-2}$. The transitional column densities are similar to that of the five sub-regions at $(0.72-1.71)\times 10^{22}\,\rm cm^{-2}$, and of the whole N-PDF at $1.62\times 10^{22}\,\rm cm^{-2}$, showing the stability of the transitional points at different scales.

\section{Analysis}\label{sec:analysis}

\subsection{N-PDF and YSOs}\label{sec:YSO}

We compare our data with the YSO data from \citet{Kuhn2021} who identified the Class-I, II, and III YSOs in Cygnus-X. The YSOs were identified using Spitzer/IRAC data \citep{Benjamin2003,Churchwell2009,Beerer2010}. The YSOs are marked in Figure \ref{fig:nmap_32}. For our analysis, we focused on YSOs with column densities above 2.64$\times 10^{21}$ $\rm cm^{-2}$, which corresponds to the lower limit used in our N-PDF analysis. In addition to the general properties of all YSOs, we also examined the relations between the N-PDF parameters and the properties of Class-I YSOs (the youngest population identified in \citet{Kuhn2021}). The external regions of Class-I YSOs still retain a significant amount of gas, providing a snapshot of the current star formation activity level, without the influence of more mature YSOs, which have already cleared much of their surrounding material.

\begin{figure}
\centering
\includegraphics[width=\figwidth]{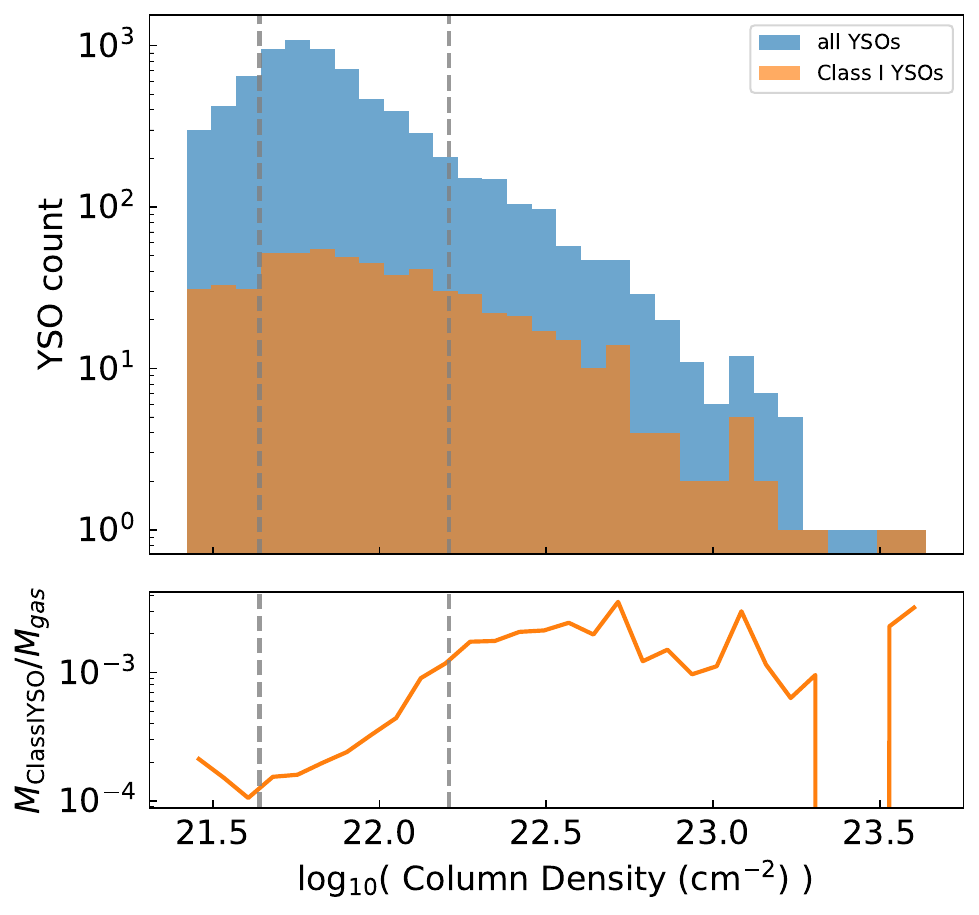}
\caption{\textit{The upper panel:} histograms of the column densities corresponding to the YSOs' positions. \textit{The lower panel:} the Class-I YSO to gas mass ratio at different column densities. The dashed lines in both panels show the $N_{\rm peak}$ and $N_{\rm TP}$ of Cygnus-X. }
\label{fig:YSOhist}
\end{figure}

The histogram in Figure \ref{fig:YSOhist} shows the distribution of column densities corresponding to the YSOs' positions. The peaks of the distribution for both `all YSOs' and the `Class-I YSOs' are at $5.25 \times 10^{21} \text{ cm}^{-2}$, close to the $N_{\rm peak}$ of Cygnus-X's log-normal part of N-PDF. The distribution of Class-I concentrates at a higher column density than that of all YSOs, consistent with the idea that younger YSOs are more associated with the dense star-forming gas. 
Adopting $0.5M_{\odot}$ as the average mass of the YSOs \citep{Muench2007}, we plot the Class-I YSO to gas mass ratio across varying column densities in the lower panel of Figure \ref{fig:YSOhist}. 
Class-I YSOs are still embedded in their natal gas, making their spatial distribution a more direct reflection of the immediate star formation process. The ratio of Class-I YSO to gas mass serves as an indicator of the current star formation efficiency at various column densities. This ratio's trend can help identify the star-forming gas. At the lowest density, the ratio is as low as $10^{-4}$, indicating inefficient star formation. However, as the column density approaches and exceeds the transitional column density at $1.62\times 10^{22}$ $\rm cm^{-2}$, the ratio increases and stabilizes at approximately $10^{-3}$. This suggests that the gas above the transitional column density exhibits a stable and efficient star-forming capability. Therefore, the transitional column density can be interpreted as a star formation threshold, above which the gas is actively engaged in star formation with a consistent efficiency. 

\begin{figure}
\centering
\includegraphics[width=\figwidth]{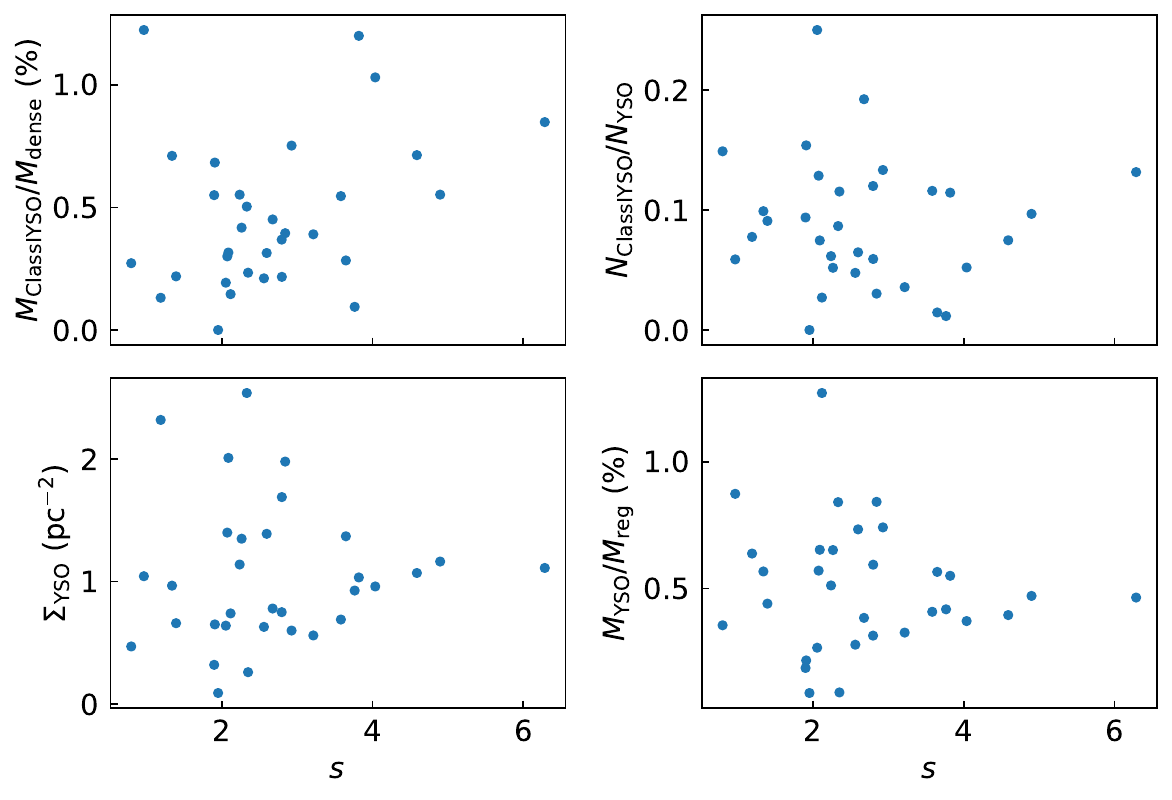}
\caption{Distributions of the Class-I YSO to dense-gas mass ratios, the Class-I YSO fractions, the YSO surface densities, and the YSO to gas mass ratios of the 32 dense regions with the power-law indices. Where $M_{\rm dense}$ is the dense-gas mass, i.e., the mass of gas above the local transitional column density, and $M_{\rm reg}$ is the total mass of the region. }
\label{fig:yso_nonrela}
\end{figure}

We further obtain the Class-I YSO to dense-gas mass ratios of the 32 dense regions, where `dense-gas' refers to the gas with column densities higher than the local transitional column density. The Class-I YSO to dense-gas mass ratio of the 32 dense regions, with the 95\% confidence interval at (0.46$\pm$0.59)\%, exhibits no correlation with the N-PDF's power-law index (see the first panel in Figure \ref{fig:yso_nonrela}). 
Additionally, the Class-I YSO fraction, YSO surface density (i.e., the normalized YSO count), and YSO to gas mass ratio also show no dependence on the power-law index. These properties are tightly distributed, with the 95\% confidence intervals at (8.79$\pm$10.43)\%, 1.04$\pm$1.12, and (0.50$\pm$0.48)\%, respectively. These results indicate that the YSO characteristics are consistent across the 32 regions. Variations in star formation, as implied by the index $s$, do not manifest in the YSO attributes of these 10 pc small sub-regions in Cygnus-X.

\subsection{N-PDF properties and star formation}

\begin{figure}
    \centering
    \includegraphics[width=0.9\figwidth]{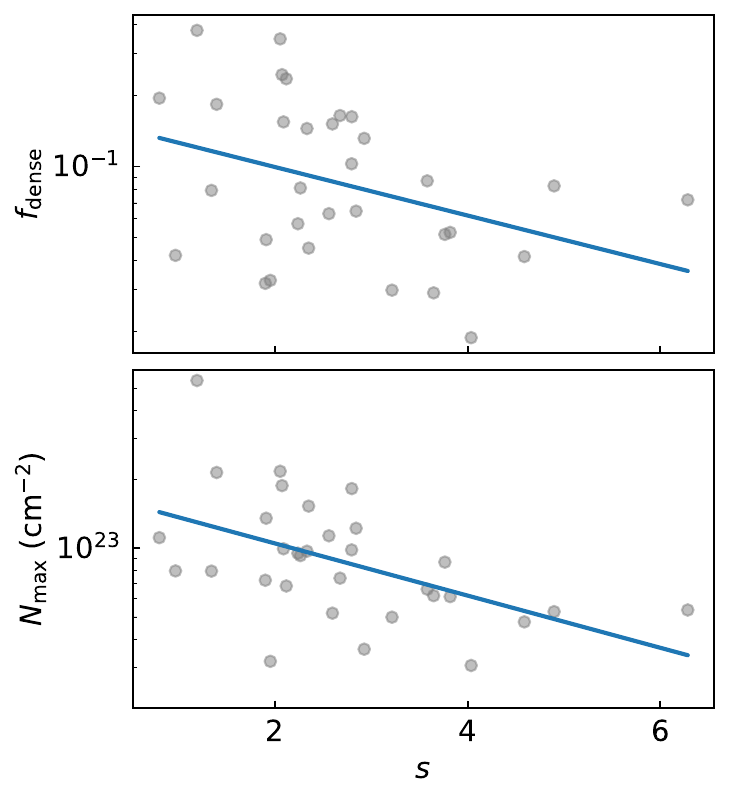}
    \caption{\textit{The upper panel: }The dense-gas mass fraction - power-law index relation of the 32 dense regions. \textit{The lower panel: }The maximum column density - power-law index relation of the 32 dense regions. The fitting results of both panels are shown in blue. }
    \label{fig:Mtp_slope}
\end{figure}

For the 32 dense regions in Cygnus-X with a scale of 10 pc, we find a correlation between the power-law index and the dense-gas mass fraction, as depicted in Figure \ref{fig:Mtp_slope}. Specifically, we observe that the dense-gas mass fraction decreases with an increasing power-law index, which can be expressed by $\log _{10} f_{\rm dense}=\log _{10}0.16-0.10s$. The relation has a Pearson correlation coefficient of 0.36, indicating a weak correlation. 
Additionally, we find that the maximum column density of each region is negatively correlated with the power-law index, with the fitting result given by $\log _{10} N_{\rm max}=\log _{10}(1.74\times 10^{23})-0.11s$, with a Pearson coefficient of 0.51.
Considering that during gravitational collapse, the amount of gravity-bounded gas and the maximum column density of the region is expected to increase, and that a shallower power-law index corresponds to a steeper density profile. These two relations may suggest that the density profile of the collapsing gas grows steeper over time, consistent with simulations of collapsing cores \citep{Kritsuk2011, Collins2012, Ward2014}. For the regions in Cygnus-X, the lowest N-PDF power-law index corresponds to a dense-gas mass fraction of approximately 10$\%$, and a maximum column density of $10^{23} \text{ cm}^{-2}$.

\begin{figure}
    \centering
    \includegraphics[width=0.8\figwidth]{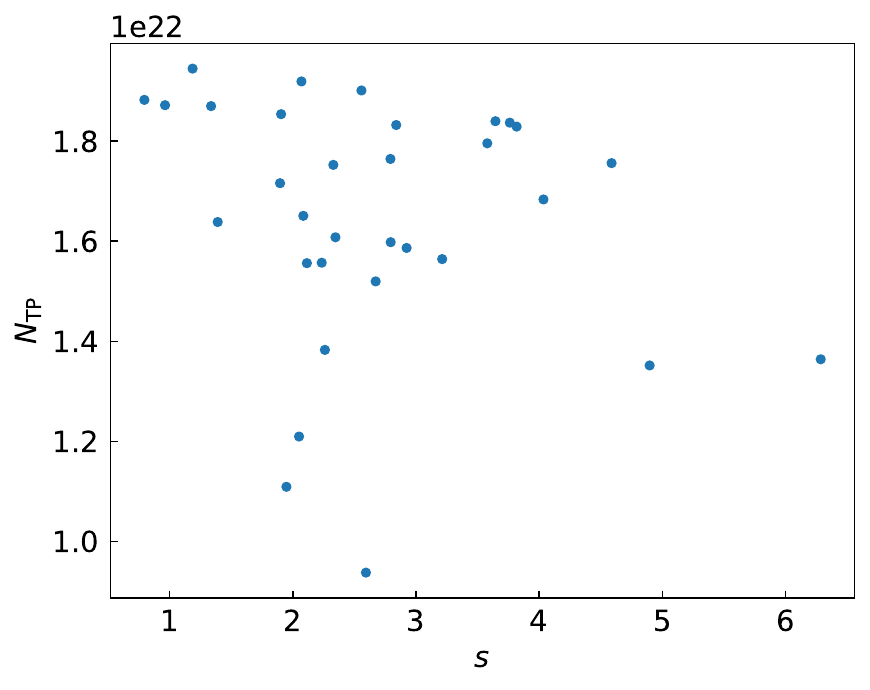}
    \caption{The transitional column density - power-law index distributions of the 32 dense regions. } 
    \label{fig:tp_slope}
\end{figure}

The transitional column densities and the power-law indices of the 32 dense regions are shown in Figure \ref{fig:tp_slope}. Visually, one may think that the transitional column densities of the 32 dense regions have a vague downward trend with the increase of the power-law indices, but the Pearson coefficient of the fitting result is less than $0.2$. Although transitional column densities with a variety of values from $10^{21}\,\rm cm^{-2}$ to $3\times 10^{22}\,\rm cm^{-2}$ have been observed from different molecular clouds \citep{Russeil2013, Schneider2013}, those of the 32 dense regions have a tight distribution, with a $95\%$ confidence interval of $(1.67\pm0.44)\times 10^{22}\,\rm cm^{-2}$. As in Table \ref{tab:wholeand5}, the transitional column densities of the whole Cygnus-X and the sub-regions also mainly concentrate at around $(1-2)\times 10^{22}\,\rm cm^{-2}$. It shows that the transitional column density is a stable parameter of a giant molecular cloud. Despite the different power-law indices, as well as a variety of maximum column densities and transitional column densities, these regions with scales of $10\,\rm pc$ in Cygnus-X have similar transitional column densities.

\subsection{Dense structures defined by the transition point}\label{sec:TPclump}

Taking the whole N-PDF transition point $1.62\times 10^{22} \,\rm cm^{-2}$ as the column density threshold, we obtain 219 dense structures as in Figure \ref{fig:nmap}. All interconnected pixels with column density larger than $1.62\times 10^{22} \,\rm cm^{-2}$ are considered as one dense structure. Structures with diameters smaller than the resolution $0.12\,\rm pc$ are ignored in this process. 

\begin{figure}
    \centering
    \includegraphics[width=\figwidth]{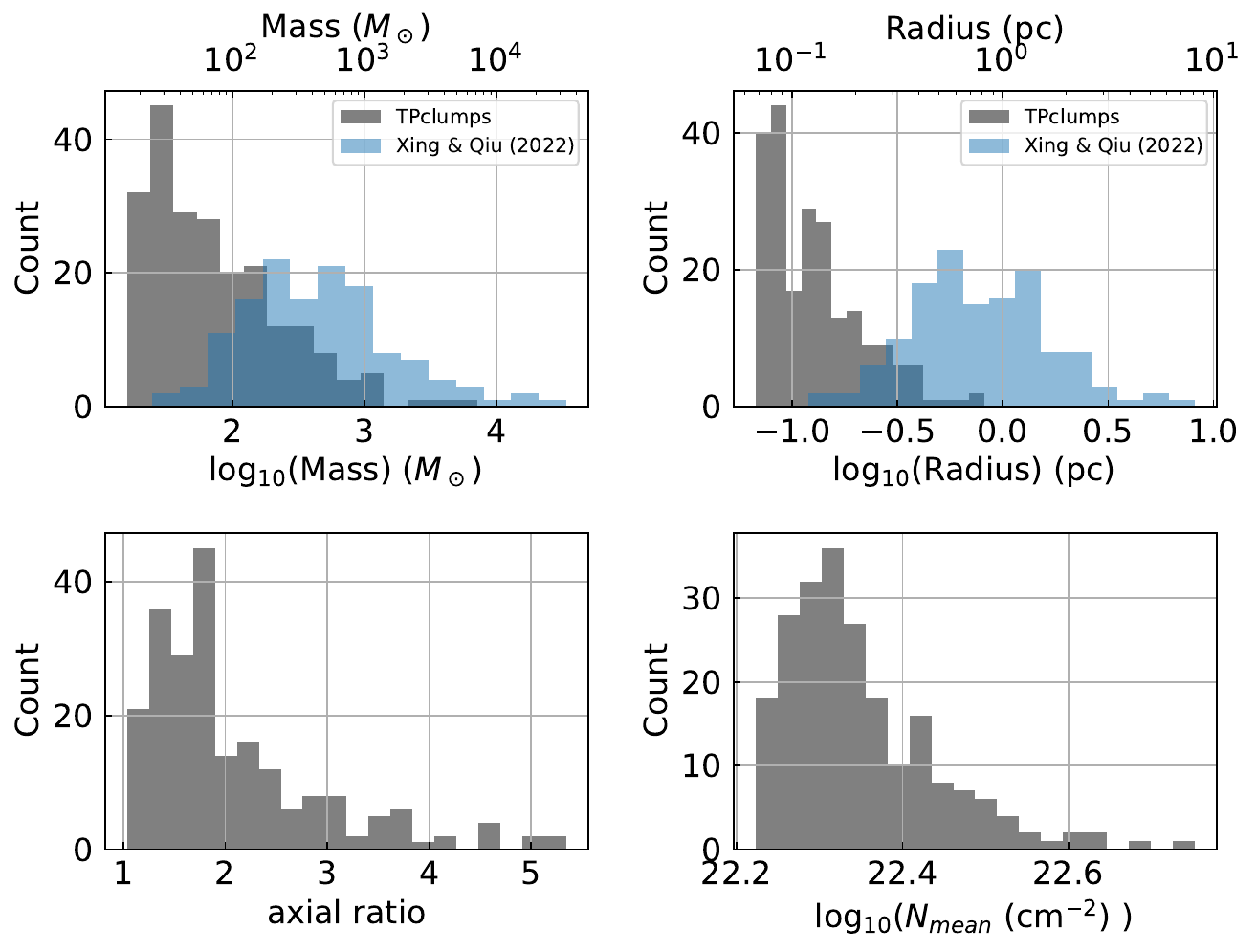}
    \caption{The grey histograms show the radii, masses, axial ratios, mean column densities, and mean number densities of the 219 dense structures. The blue histograms in the upper panels show the radii and masses of the 135 structures' inner parts in \citet{Xing2022}.
    }
    \label{fig:TPclump}
\end{figure}

The structures make up the dense component of Cygnus-X, thus, they reflect the properties of the gravity-dominated gas. 
We plot the distributions of the radii, masses, axial ratios, mean column densities, and mean number densities of the 219 structures in Figure \ref{fig:TPclump}. The effective radii are the geometric means between the FWHMs of the major and minor axes of the clumps. The dense structures have effective radii from $0.82\,\rm pc$ down to the resolution limit at $R=0.06\,\rm pc$. The masses of dense structures vary from $15.8\,\rm M_\odot$ to $7099.2\,\rm M_\odot$, covering three orders of magnitude. Dense structures with smaller sizes and masses are always more numerous than those with larger ones. The ratios of major to minor axes of dense structures range from $1.03$ to $5.35$. $61.6\%$ of the ratios are between 1 and 2, which means that the projections of the dense structures generally have circular morphology. Assuming the spatial inclinations of the dense structures have a uniform distribution, the ratios suggest that the dense structures are mostly spherical, proving the reliability of our spherical assumption for the gravity-dominated part. The mean column densities of the dense structures range from $1.67\times 10^{22}\,\rm cm^{-2}$ to $5.66\times 10^{22}\,\rm cm^{-2}$. The distribution has a peak at $2\times 10^{22} \,\rm cm^{-2}$, which is five times higher than the column density of the turbulence-dominated part in Cygnus-X. 

Using the same Cygnus-X column density map, \citet{Xing2022} obtained and fitted radial profiles of 135 dense clumps in Cygnus-X. The clumps have broken power-law column density
profiles, whose parameters are consistent with the gravity/turbulence dominated circumstance at the inner/outer part. We expect the inner part of the dense clumps in \citet{Xing2022} to be comparable to our dense structures. However, as in Figure \ref{fig:TPclump}, the properties of these two sets of structures are quite different. The radii and masses of \citet{Xing2022} clumps do not have bottom-heavy distributions but have peaks at 0.8 pc\footnote{Note that the radii values of the two set of structures cannot be compared, because the radii in \citet{Xing2022} are not effective radii.} and $4.5\times 10^2$ $M_{\odot}$. We think this difference arises from the bias introduced by using a single column density threshold to select the dense structures. Our result in Section \ref{sec:5reg} shows that the sub-regions of Cygnus-X have different transition column densities. When a uniform density threshold is applied to select dense structures, some small structures that are primarily turbulence-dominated are inevitably included. These structures have small sizes and low densities, and they constitute the lower, heavy portion of Figure \ref{fig:TPclump}. Therefore, when extracting information about the gravity-dominated component of dense structures, those with larger masses are more relevant.

\begin{figure}
    \centering
    \includegraphics[width=\figwidth]{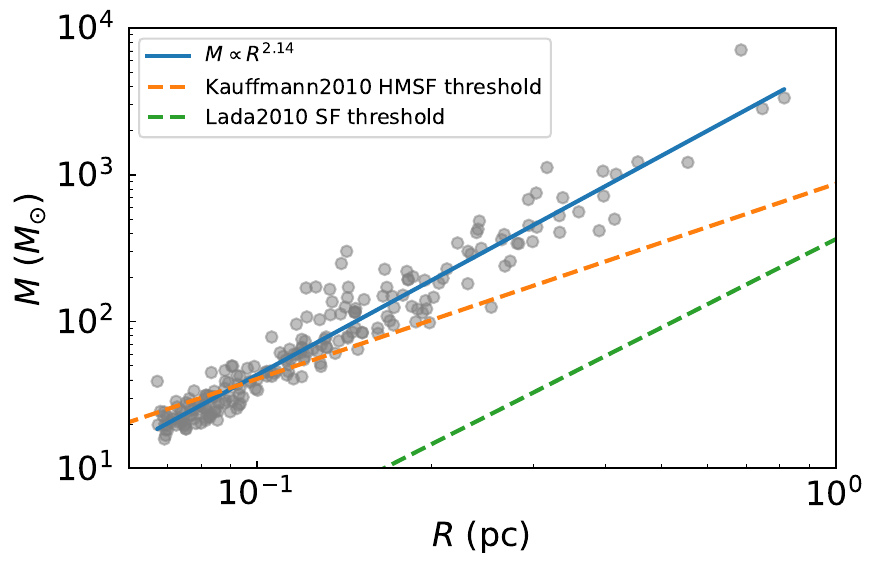}
    \caption{The mass-radius relation of the dense structures. The blue solid line shows the fitting result. The dashed lines show star formation thresholds as in the labels. }
    \label{fig:MR}
\end{figure}

We plot the mass - size relation of the 219 dense structures in Figure \ref{fig:MR}. The fitting result of the relation is $M\propto R^{2.14}$, which indicates that the radial density indices of the dense structures are slightly different \citep{Xing2022}. All the dense structures have masses higher than the \citet{Lada2010} empirical star-formation threshold, verifying their star formation potential. 136 out of the 219 dense structures satisfy the empirical high-mass star formation threshold proposed by \citet{Kauffmann2010_threshold}. Those dense structures that do not meet the threshold, however, are also very close to the threshold. It shows that the dense structures may be able to form massive stars.

\section{Discussion}\label{sec:discuss}

\subsection{N-PDFs as accessible statistical tools}

Since \citet{Kainulainen2009}, N-PDFs derived from dust extinction and continuum data have become one of the most preferred tools for characterizing molecular clouds. As a statistical measurement, N-PDFs offer an intuitive glimpse into the internal density distribution of the clouds. Compared to other methods for obtaining the density distribution of molecular clouds, such as radial density profiles, N-PDFs are more readily accessible. They are simple histograms of density distributions, eschewing the need for structure identification algorithms and thereby avoiding the selection biases. Moreover, N-PDFs are not contingent upon high-resolution data, which is a significant advantage in scenarios where observational conditions are limited or where high-resolution data acquisition is challenging or costly. This characteristic is particularly valuable for studying distant or observationally intractable molecular clouds. The statistical approach of N-PDFs is also unaffected by local region selection, providing statistical properties across the entire observed area. This capability is especially useful for obtaining the density threshold where gravity begins to dominate, as the transition point is easy to find. On the contrary, discerning the boundary between dense cores and the diffuse environment has been challenging in radial density profile analysis.

\subsection{N-PDF transitions as star formation thresholds}

The transition point in the N-PDF divides areas dominated by turbulence and gravity. Theoretically, it is natural to consider this transition as a threshold for star formation. However, we also know that the N-PDF, as a statistical quantity and subject to projection effects, the right side of its transition point should not be simply equated to regions governed by gravity. Consequently, whether the transition point can be regarded as a star formation threshold or if a function of the point can serve as a star formation threshold has remained an important question. In Section \ref{sec:analysis}, we find that the transition point is closely correlated with the mass fraction of Class-I YSOs. Above the transitional column density, the ratio of Class-I YSO to gas mass remains stable at approximately $10^{-3}$, while to the left of the transition point, this ratio can be an order of magnitude lower. Given that this ratio reflects the current star formation efficiency at different densities, this indicates that regions above the transition point are star-forming gas with a consistent efficiency. This supports the interpretation of the transition point as a star formation threshold, where the gas transitions from inefficient to efficient star-forming states.

For a molecular cloud, the star formation threshold represented by the transition point is stable. In the case of Cygnus-X, we find that all N-PDFs we get, ranging from that of the entire cloud to sub-regions as small as 10 pc, share similar transition points. The N-PDF power-law indices, dense gas mass fractions, and maximum column densities of the sub-regions differ significantly, their relations are consistent with models of collapsing cores, suggesting that these regions may be at different stages of gravitational collapse. Despite these variations, the transition points for all regions cluster around $1.67 \times 10^{22}\, \rm cm^{-2}$. Additionally, the YSO characteristics across these regions are similar and unrelated to the N-PDF power-law indices. This indicates that, despite differences in evolutionary age and physical conditions between sub-regions, the regions within the same molecular cloud may still share similar star formation thresholds and efficiencies. The overall properties of the molecular cloud seem to play a more decisive role in determining the star formation threshold. Whether considering local or global properties of the cloud, or different evolutionary stages, the star formation threshold remains stable. Conversely, the N-PDF power-law indices are more sensitive to time and can thus be used as tracers of the evolutionary stages of molecular clouds and their internal regions.

Additionally, it is important to note that the transition point only represents a star formation threshold that is specific to the current molecular cloud and not a universal threshold. Due to variations in the radiation field and thickness of each molecular cloud, the actual density threshold for gravitationally bound gas of each molecular cloud can be different \citep{Clark2013}. Observations have also shown that the N-PDF transition point can vary by an order of magnitude between different molecular clouds \citep{Schneider2022}. Taking the empirical star formation threshold of $7.3\times10^{21}$ $\rm cm^{-2}$ ($A_{\rm K}$=0.8 mag) derived by \citet{Lada2010} as an example, we find no significant difference in the mass fraction of Class-I YSOs on either side of this column density in Cygnus-X. This threshold, although applicable to a lot of nearby molecular clouds, does not apply to Cygnus-X.

\section{Conclusion}\label{sec:sum}

As part of the CENSUS project, we use the column density map from \citet{Cao2019} to obtain the N-PDFs of Cygnus-X, providing a detailed statistical analysis of the cloud's density distribution and their relation to star formation. The main results are summarized as follows: 

\begin{enumerate}
    \item Log-normal + Power-law N-PDF Shape: The N-PDFs of Cygnus-X and four out of five of its sub-regions exhibit log-normal + power-law shapes, indicative of the combined effects of turbulence and gravity. It is consistent with previous studies of star-forming molecular clouds, confirming the universality of this density distribution pattern.
    \item HI N-PDF: The Cygnus‑X HI N-PDF shows a log‑normal‑like shape but is skewed toward higher column densities. It is much narrower than the H\(_2\) N‑PDF, with most HI confined to \(N_{\rm HI}\sim4\text{–}8\times10^{21}\,\mathrm{cm^{-2}}\) and peaking at \(5.5\times10^{21}\,\mathrm{cm^{-2}}\). Both the peak and the high‑density truncation of the HI N-PDF lie below the H\(_2\) N‑PDF peak, suggesting a link between the HI N-PDF morphology and the HI‑to‑H$_2$ transition.
    \item Star Formation Threshold: With YSO data, we see that the transitional column density in Cygnus-X, at $1.62 \times 10^{22}$ cm$^{-2}$, serves as a star formation threshold for Cygnus-X. Above the threshold, the ratio of Class-I YSO to gas mass remains stable at approximately $10^{-3}$, exhibiting a uniform star formation efficiency.
    \item Evolution of N-PDF Parameters: We see linear relations between the power-law index, the dense-gas fraction, and the local maximum column density. They suggest that the power-law segment of the N-PDFs flattens over time, suggesting an evolution in the density structure of the cloud. While the transitional column density, which separates the log-normal and power-law regimes, remains the same, suggesting that the overall properties of the molecular cloud play a more significant role in determining this star formation threshold than local variations.
    \item N-PDF as an accessible Tool: N-PDFs are a simple and accessible way to characterize the density distribution of molecular clouds, without the need for complex structure identification algorithms or high-resolution data. By obtaining key statistical properties such as the transition point and power-law index, they can serve as powerful tools in studies of star-forming regions.
\end{enumerate}

\begin{acknowledgments}
This work is supported by National Natural Science Foundation of China (NSFC) grants Nos. 12425304 and U1731237, the National Key R\&D Program of China with Nos. 2023YFA1608204 and 2022YFA1603103, and the grant from the China Manned Space Project.
\end{acknowledgments}


\bibliography{bibtex}{}
\bibliographystyle{aasjournal}

\end{document}